\documentclass[onecolumn,prd,nofootinbib,aps,floats,floatfix,amsmath,amssymb,secnumarabic,preprintnumbers,showpacs,notitlepage]{revtex4-1} %
\usepackage[final]{graphicx}

\def\beq{\begin{equation}}
\def\eeq{\end{equation}}
\def\bea{\begin{eqnarray}}
\def\eea{\end{eqnarray}}
\def\nn{\nonumber}
\def\roughly#1{\mathrel{\raise.3ex\hbox{$#1$\kern-.75em\lower1ex\hbox{$\sim$}}}}
\def\lsim{\roughly<}
\def\gsim{\roughly>}

\def\bs{B^0_s}
\def\bsbar{{\bar B}^0_s}
\def\bsmumu{b \to s \mu^+ \mu^-}
\def\bsnunubar{b \to s \nu {\bar\nu}}

\def\BKstarmumu{B \to K^* \mu^+ \mu^-}
\def\BKmumu{B \to K \mu^+ \mu^-}

\def \SM{{\rm SM}}

\def \expt{{\rm expt}}

\def\s{\sqrt{2}}
\def\bsll{b \to s \ell^+ \ell^-}

\newcommand{\av}[1]{\langle #1 \rangle}

\def\gsim{{~\raise.15em\hbox{$>$}\kern-.85em
          \lower.35em\hbox{$\sim$}~}}
\def\lsim{{~\raise.15em\hbox{$<$}\kern-.85em
          \lower.35em\hbox{$\sim$}~}}

\newcommand{\gev}{\ensuremath{\mathrm{\,Ge\kern -0.1em V}}}

\begin{document}

\title{\boldmath New Physics in $\bsmumu$: Distinguishing Models through CP-Violating Effects}
\preprint{UdeM-GPP-TH-17-255; WSU-HEP-1703}
\author{Ashutosh Kumar Alok}
\email{akalok@iitj.ac.in}
\affiliation{Indian Institute of Technology Jodhpur, Jodhpur 342011, India}
\author{Bhubanjyoti Bhattacharya}\email{bhujyo@wayne.edu}
\affiliation{Department of Physics and Astronomy, \\
Wayne State University, Detroit, MI 48201, USA}
\author{Dinesh Kumar}\email{dinesh@phy.iitb.ac.in}
\affiliation{Indian Institute of Technology Bombay, Mumbai 400076, India}
\affiliation{Department of Physics, University of Rajasthan, Jaipur 302004, India}
\author{Jacky Kumar}\email{jka@tifr.res.in}
\affiliation{Department of High Energy Physics, Tata Institute of Fundamental Research, \\
400 005, Mumbai, India}
\author{David London}\email{london@lps.umontreal.ca}
\affiliation{Physique des Particules, Universit\'e de Montr\'eal, \\
C.P. 6128, succ. centre-ville, Montr\'eal, QC, Canada H3C 3J7}
\author{S. Uma Sankar}\email{uma@phy.iitb.ac.in}
\affiliation{Indian Institute of Technology Bombay, Mumbai 400076, India}

\begin{abstract}
At present, there are several measurements of $B$ decays that exhibit
discrepancies with the predictions of the SM, and suggest the presence
of new physics (NP) in $\bsmumu$ transitions. Many NP models have been
proposed as explanations.  These involve the tree-level exchange of a
leptoquark (LQ) or a flavor-changing  $Z'$ boson. In this paper we
examine whether it is possible to distinguish the various models via
CP-violating effects in $B\to K^{(*)}\mu^+\mu^-$. Using fits to the
data, we find the following results. Of all possible LQ models, only
three can explain the data, and these are all equivalent as far as
$\bsmumu$ processes are concerned. In this single LQ model, the weak
phase of the coupling can be large, leading to some sizeable CP
asymmetries in $B\to K^{(*)}\mu^+\mu^-$. There is a spectrum of $Z'$
models; the key parameter is $g_L^{\mu\mu}$, which describes the
strength of the $Z'$ coupling to $\mu^+\mu^-$. If $g_L^{\mu\mu}$ is
small (large), the constraints from $\bs$-$\bsbar$ mixing are stringent
(weak), leading to a small (large) value of the NP weak phase, and
corresponding small (large) CP asymmetries. We therefore find that the
measurement of CP-violating asymmetries in $B\to K^{(*)}\mu^+\mu^-$ can
indeed distinguish among NP $\bsmumu$ models.
\end{abstract}


\maketitle

\section{Introduction}
\label{Sec:Intro}

At present, there are several measurements of $B$ decays involving
$\bsll$ that suggest the presence of physics beyond the standard model
(SM). These include
\begin{enumerate}

\item $B \to K^* \mu^+\mu^-$: Measurements of $B \to K^* \mu^+\mu^-$
  have been made by the LHCb \cite{BK*mumuLHCb1,BK*mumuLHCb2} and
  Belle \cite{BK*mumuBelle} Collaborations. They find results that
  deviate from the SM predictions. The main discrepancy is in the
  angular observable $P'_5$ \cite{P'5}. Its significance depends on
  the assumptions made regarding the theoretical hadronic
  uncertainties
  \cite{BK*mumuhadunc1,BK*mumuhadunc2,BK*mumuhadunc3}. The latest fits
  to the data \cite{Altmannshofer:2014rta,BK*mumulatestfit1,BK*mumulatestfit2} take into
  account the hadronic uncertainties, and find that a significant
  discrepancy is still present, perhaps as large as $\sim 4\sigma$.

\item $\bs \to \phi \mu^+ \mu^-$: The LHCb Collaboration has
  measured the branching fraction and performed an angular analysis of
  $\bs \to \phi \mu^+ \mu^-$
  \cite{BsphimumuLHCb1,BsphimumuLHCb2}. They find a $3.5\sigma$
  disagreement with the predictions of the SM, which are based on
  lattice QCD \cite{latticeQCD1,latticeQCD2} and QCD sum rules
  \cite{QCDsumrules}.

\item $R_K$: The ratio $R_K \equiv {\cal B}(B^+ \to K^+ \mu^+
  \mu^-)/{\cal B}(B^+ \to K^+ e^+ e^-)$ has been measured by the LHCb
  Collaboration in the dilepton invariant mass-squared range 1 GeV$^2$
  $\le q^2 \le 6$ GeV$^2$ \cite{RKexpt}, with the result
\bea
R_K^\expt = 0.745^{+0.090}_{-0.074}~{\rm (stat)} \pm 0.036~{\rm (syst)} ~.
\label{RKexp}
\eea
This differs from the SM prediction of $R_K^\SM = 1 \pm 0.01$
\cite{IsidoriRK} by $2.6\sigma$, and thus is a hint of lepton flavor
non-universality.

\end{enumerate}

While any suggestions of new physics (NP) are interesting, what is
particularly intriguing about the above set of measurements is that
they can all be explained if there is NP in $\bsmumu$\footnote{Early
  model-independent analyses of NP in $\bsmumu$ can be found in
  Refs.~\cite{bsmumuCPC} (CP-conserving observables) and
  \cite{bsmumuCPV} (CP-violating observables).}. To be specific,
$\bsmumu$ transitions are defined via the effective Hamiltonian
\bea
H_{\rm eff} &=& - \frac{\alpha G_F}{\s \pi} V_{tb} V_{ts}^*
      \sum_{a = 9,10} ( C_a O_a + C'_a O'_a ) ~, \nn\\
O_{9(10)} &=& [ {\bar s} \gamma_\mu P_L b ] [ {\bar\mu} \gamma^\mu (\gamma_5) \mu ] ~,
\label{Heff}
\eea
where the $V_{ij}$ are elements of the Cabibbo-Kobayashi-Maskawa (CKM)
matrix. The primed operators are obtained by replacing $L$ with $R$,
and the Wilson coefficients (WCs) $C^{(\prime)}_a$ include both SM and
NP contributions. Global analyses of the $\bsll$ anomalies have been
performed
\cite{Descotes-Genon:2013wba,Altmannshofer:2014rta,BK*mumulatestfit1,BK*mumulatestfit2}. It
was found that there is a significant disagreement with the SM,
possibly as large as $4\sigma$, and it can be explained if there is NP
in $b \to s \mu^+ \mu^-$. Ref.~\cite{BK*mumulatestfit1} gave four
possible explanations: (I) $C_9^{\mu\mu}({\rm NP}) < 0$, (II)
$C_9^{\mu\mu}({\rm NP}) = - C_{10}^{\mu\mu}({\rm NP}) < 0$, (III)
$C_9^{\mu\mu}({\rm NP}) = - C_{9}^{\prime \mu\mu}({\rm NP}) < 0$, (IV)
$C_9^{\mu\mu}({\rm NP}) = - C_{10}^{\mu\mu}({\rm NP}) = -C_{9}^{\prime
  \mu\mu}({\rm NP}) = - C_{10}^{\prime \mu\mu}({\rm NP}) < 0$.

Numerous models have been proposed that generate the correct NP
contribution to $\bsmumu$ at tree level\footnote{The anomalies can
  also be explained using a scenario in which the NP enters in the $b
  \to c {\bar c} s$ transition, but constraints from radiative $B$
  decays and $\bs$-$\bsbar$ mixing must be taken into account, see
  Ref.~\cite{AlexLenz}.}. Most of them use solution (II) above, though
a few use solution (I). These models can be separated into two
categories: those containing leptoquarks (LQs)
\cite{CCO,AGC,HS1,GNR,VH,SM,FK,BFK,BKSZ}, and those with a $Z'$ boson
\cite{CCO,Crivellin:2015lwa,Isidori,dark,Chiang,Virto,GGH,BG,BFG,Perimeter,CDH,SSV,CHMNPR,CMJS,BDW,FNZ,AQSS,CFL,Hou,CHV,CFV,CFGI,IGG,BdecaysDM,Bhatia:2017tgo}. But
this raises an obvious question: assuming that there is indeed NP in
$\bsmumu$, which model is the correct one? In other words, short of
producing an actual LQ or $Z'$ experimentally, is there any way of
distinguishing the models?

A first step was taken in Ref.~\cite{RKRDmodels}, where it was shown
that the CP-conserving, lepton-flavor-violating decays $\Upsilon(3S)
\to \mu \tau$ and $\tau \to 3\mu$ are useful processes for
differentiating between LQ and $Z'$ models. In the present paper, we
compare the predictions of the various models for CP-violating
asymmetries in $\BKstarmumu$ and $\BKmumu$.

CP-violating effects require the interference of two amplitudes with a
relative weak (CP-odd) phase. (For certain CP-violating effects, a
relative strong (CP-even) phase is also required.) In the SM,
$\bsmumu$ is dominated by a single amplitude, proportional to $V_{tb}
V_{ts}^*$ [see Eq.~(\ref{Heff})]. In order to generate CP-violating
asymmetries, it is necessary that the NP contribution to $\bsmumu$
have a sizeable weak phase. As we will see, this does not hold in all
NP models, so that CP-violating asymmetries in $\BKstarmumu$ and
$\BKmumu$ can be a powerful tool for distinguishing the models. (The
usefulness of CP asymmetries in $\BKstarmumu$ for identifying NP was
also discussed in Ref.~\cite{BHP}.)

We perform both model-independent and model-dependent analyses. In the
model-independent case, we assume that the NP contributes to a
particular set of WCs (and we consider several different sets). But if
a particular model is used, one can work out which WCs are affected.
In either case, a fit to the data is performed to establish (i)
whether a good fit is obtained, and (ii) what are the best-fit values
and allowed ranges of the real and imaginary pieces of the WCs. In the
case of a good fit, the predictions for CP-violating asymmetries in
$\BKstarmumu$ and $\BKmumu$ are computed.

The data used in the fits include all CP-conserving observables
involving $\bsmumu$ transitions. The processes are $B^0 \to K^{*0}
(\to K^+ \pi^-) \mu^+ \mu^-$, $B^+ \to K^{*+} \mu^+ \mu^-$, $B^+ \to
K^+ \mu^+ \mu^-$, $B^0 \to K^0 \mu^+ \mu^-$, $\bs \to \phi \mu^+
\mu^-$, $B \to X_s \mu^+ \mu^-$, and $\bs \to \mu^+ \mu^-$. For the
first process, a complete angular analysis of $B^0 \to K^{*0} (\to K^+
\pi^-) \mu^+ \mu^-$ was performed in Refs.~\cite{BHP,BK*mumuCPV}. It
was shown that this decay is completely described in terms of twelve
angular functions. By averaging over the angular distributions of $B$
and ${\bar B}$ decays, one obtains CP-conserving observables. There
are nine of these. Most of the observables are measured in different
$q^2$ bins, so that there are a total of 106 CP-conserving observables
in the fit.

For the model-independent fits, only the $\bsmumu$ data is used.
However, for the model-dependent analyses, additional data may be
taken into account. That is, in a specific model, there may be
contributions to other processes such as $\bsnunubar$, $\bs$-$\bsbar$
mixing, etc. The choice of additional data is made on a model-by-model
basis. Because the model-independent and model-dependent fits can
involve different experimental (and theoretical) constraints, they may
yield significantly different results.

CP-violating asymmetries are obtained by comparing $B$ and ${\bar B}$
decays. In the case of $B \to K \mu^+ \mu^-$, there is only the direct
partial rate asymmetry. For $B^0 \to K^{*0} (\to K^+ \pi^-) \mu^+
\mu^-$, one compares the $B$ and ${\bar B}$ angular distributions.
This leads to seven CP asymmetries. There are therefore a total of
eight CP-violating effects that can potentially be used to distinguish
among the NP $\bsmumu$ models.

For the LQs, we will show that there are three models that can explain
the $\bsmumu$ data. The LQs of these models contribute differently to
$b \to s \nu_\mu {\bar\nu}_\mu$, so that, in principle, they can be
distinguished by the measurements of $\bsnunubar$. However, the
constraints from these measurements are far weaker than those from
$\bsmumu$, so that all three LQ models are equivalent, as far as the
$\bsmumu$ data are concerned.  We find that some CP asymmetries in $B
\to K^{(*)} \mu^+ \mu^-$ can be large in this single LQ model.

In $Z'$ models, there are $g_L^{bs} {\bar s} \gamma^\mu P_L b Z'_\mu$
and $g_L^{\mu \mu} {\bar \mu} \gamma^\mu P_L \mu Z'_\mu$ couplings,
leading to a tree-level $Z'$ contribution to $\bsmumu$. In order to
explain the $\bsmumu$ anomalies, the product of couplings $g_L^{bs}
g_L^{\mu \mu}$ must lie within a certain (non-zero) range. If
$g_L^{\mu \mu}$ is small, $g_L^{bs}$ must be large, and vice-versa.
The $Z'$ also contributes at tree level to $\bs$-$\bsbar$ mixing,
proportional to $(g_L^{bs})^2$. Measurements of the mixing constrain
the magnitude and phase of $g_L^{bs}$. If $g_L^{bs}$ is large, the
constraint on its phase is significant, so that this $Z'$ model cannot
generate sizeable CP asymmetries. On the other hand, if $g_L^{bs}$ is
small, the constraints from $\bs$-$\bsbar$ mixing are not stringent,
and large CP-violating effects are possible.

The upshot is that it may be possible to differentiate $Z'$ and LQ
models, as well as different $Z'$ models, through measurements of
CP-violating asymmetries in $B \to K^{(*)} \mu^+ \mu^-$.

We begin in Sec.~2 with a description of our method for fitting the
data and for making predictions about CP asymmetries. The $\bsmumu$
data used in the fits are given in the Appendix. We perform a model%
-independent analysis in Sec.~3. In Sec.~4, we perform model-dependent
fits in order to determine the general features of the LQ and $Z'$
models that can explain the $\bsmumu$ anomalies. We present the
predictions of the various models for the CP asymmetries in Sec.~5.
We conclude in Sec.~6.

\section{Method}

The method works as follows. We suppose that the NP contributes to a
particular set of $\bsmumu$ WCs. This can be done in a
``model-independent'' way, in the sense that no particular underlying
NP model is assumed, or it can be done in the context of a specific NP
model. In either case, all observables are written as functions of the
WCs, which contain both SM and NP contributions.  Given values of the
WCs, we use {\tt flavio} \cite{flavio} to calculate the observables.
By comparing the computed values of the observables with the data, the
$\chi^2$ can be found.  The program {\tt MINUIT} \cite{James:1975dr,
  James:2004xla, James:1994vla} is used to find the values of the WCs
that minimize the $\chi^2$. It is then possible to determine whether
or not the chosen set of WCs provides a good fit to the data. This is
repeated for different sets of $\bsmumu$ WCs.

We are interested in NP that leads to CP-violating effects in $B \to
K^{(*)} \mu^+ \mu^-$. As noted in the introduction, this requires that
the NP contribution to $\bsmumu$ have a weak phase. With this in mind,
we allow the NP WCs to be complex (other fits generally take the NP
contributions to the WCs to be real), and determine the best-fit
values of both the real and imaginary parts of the WCs.

In the case where a particular NP model is assumed, the main
theoretical parameters are the couplings of the NP particles to the SM
fermions. At low energies, these generate four-fermion operators.  The
first step is therefore to determine which operators are generated in
the NP model. This in turn establishes which observables are affected
by the NP. The fit yields preferred values of the WCs, and these can
be converted into preferred values for the real and imaginary parts
of the couplings.

We note that caution is needed as regards the results of the
model-independent fits. In such fits it is assumed that the NP
contributes to a particular set of WCs. One might think that the
results will apply to all NP models that contribute to the same WCs.
However, this is not true. The point is that a particular model may
have additional theoretical or experimental constraints. When these
are taken into account, the result of the fit might be quite
different. That is, the ``model-independent'' fits do not necessarily
apply to all models.  Indeed, in the following sections we will see
several examples of this.

Finally, for those sets of WCs that provide good fits to the data, we
compute the predictions for the CP-violating asymmetries in
$\BKstarmumu$ and $\BKmumu$.

\subsection{\bf Fit}

The $\chi^2$ is a function of the WCs $C_i$, and is constructed as
follows:
\beq
\chi^2(C_i) = (\mathcal{O}_{th}(C_i) -\mathcal{O}_{exp})^T \, \mathcal{C}^{-1} \,
(\mathcal{O}_{th}(C_i) -\mathcal{O}_{exp}) ~.
\eeq
Here $\mathcal{O}_{th}(C_i)$ are the theoretical predictions for the
various observables used as constraints. These predictions depend upon
the WCs.  $\mathcal{O}_{exp}$ are the the corresponding experimental
measurements.

We include all available theoretical and experimental correlations in
our fit. The total covariance matrix $\mathcal{C}$ is obtained by
adding the individual theoretical and experimental covariance
matrices, respectively $\mathcal{C}_{th}$ and $\mathcal{C}_{exp}$. The
theoretical covariance matrix is obtained by randomly generating all
input parameters and then calculating the observables for these sets
of inputs \cite{flavio}.The uncertainty is then defined by the
standard deviation of the resulting spread in the observable
values. In this way the correlations are generated among the various
observables that share some common parameters \cite{flavio}.  Note
that we have assumed $\mathcal{C}_{th}$ to be independent of the
WCs. This implies that we take the SM covariance matrix to construct
the $\chi^2$ function.  As far as experimental correlations are
concerned, these are only available (bin by bin) among the angular
observables in $B \to K^{(*)} \mu^+ \mu^-$ \cite{BK*mumuLHCb2}, and
among the angular observables in $\bs \to \phi \mu^+ \mu^-$
\cite{BsphimumuLHCb2}.

For $\chi^2$ minimization, we use the {\tt MINUIT} library
\cite{James:1975dr,James:2004xla,James:1994vla}. The errors on the
individual parameters are defined as the change in the values of the
parameters that modifies the value of the $\chi^2$ function such that
$\Delta \chi^2 = \chi^2- \chi^2_{min}=1$. However, to obtain the $68.3
\%$ and $95\%$ CL 2-parameter regions, we use $\Delta \chi^2$ equal
to 2.3 and 6.0, respectively \cite{pdg}.

The fit includes all CP-conserving $\bsmumu$ observables. These are
\begin{enumerate}

\item $B^0 \to K^{*0} \mu^+ \mu^-$: The CP-averaged differential
  angular distribution for $B^0 \to K^{*0} (\to K^+ \pi^-) \mu^+
  \mu^-$ can be derived using Refs.~\cite{P'5,BHP,BK*mumuCPV}; it is
  given by \cite{BK*mumuLHCb2}
\bea
\label{angdist}
&& \frac{1}{d(\Gamma + \overline{\Gamma})/d q^2} \frac{d^4 (\Gamma + \overline{\Gamma})}
{dq^2d\vec\Omega} = \frac{9}{32 \pi} \left[\frac34 (1
- F_L) \sin^2 \theta_{K^*} + F_L \cos^2 \theta_{K^*} \right. \\
&& \hskip3truemm
+~\frac14 (1 - F_L) \sin^2 \theta_{K^*} \cos 2\theta_\ell
- F_L \cos^2 \theta_{K^*} \cos 2\theta_\ell
+ S_3 \sin^2 \theta_{K^*} \sin^2 \theta_\ell \cos 2\phi \nn\\
&& \hskip3truemm
+~S_4 \sin 2\theta_{K^*} \sin 2\theta_\ell \cos \phi
+ S_5 \sin 2\theta_{K^*} \sin \theta_\ell \cos \phi
+ \frac{4}{3} A_{FB}\sin^2 \theta_{K^*} \cos \theta_\ell \nn\\
&& \hskip3truemm \left.
+~S_7 \sin 2\theta_{K^*} \sin \theta_\ell \sin \phi
+ S_8 \sin 2\theta_{K^*} \sin 2\theta_\ell \sin \phi
+ S_9 \sin^2 \theta_{K^*} \sin^2 \theta_\ell \sin 2\phi \right] ~. \nn
\eea
Here $q^2$ represents the invariant mass squared of the dimuon system, and
$\vec\Omega$ represents the solid angle constructed from $\theta_l, \theta_{K^*}$,
and $\phi$. There are therefore nine observables in the decay: the differential
branching ratio, $F_L$, $A_{FB}$, $S_3$, $S_4$, $S_5$, $S_7$, $S_8$ and $S_9$,
all measured in various $q^2$ bins. The experimental measurements are given in
Tables \ref{B0K*mumuBRmeas} and \ref{tab:BtoKstar} in the Appendix.

In the introduction it was mentioned that the main discrepancy with
the SM is in the angular observable $P'_5$. This is defined as \cite{P'5}
\beq
P'_5 = \frac{S_5}{\sqrt{F_L (1  - F_L)}} ~.
\eeq

\item The differential branching ratio of $B^+ \to K^{*+} \mu^+
  \mu^-$. The experimental measurements \cite{Aaij:2014pli} are given in Table
  \ref{B+K*mumuBRmeas} in the Appendix.

\item The differential branching ratio of $B^+ \to K^+ \mu^+ \mu^-$.
  The experimental measurements \cite{Aaij:2014pli} are given in Table
  \ref{B+KmumuBRmeas} in the Appendix. When integrated over $q^2$,
  this provides the numerator in $R_K \equiv {\cal B}(B^+ \to K^+
  \mu^+ \mu^-)/{\cal B}(B^+ \to K^+ e^+ e^-)$. Thus, the measurement
  of $R_K$ [Eq.~(\ref{RKexp})] is implicitly included
  here\footnote{Previous studies (Ref.~\cite{RKRDmodels} and
    references therein) have indicated that the $R_K$ anomaly can be
    accommodated side-by-side with several other anomalies in $b\to
    s\mu^+\mu^-$ if new physics only affects transitions involving
    muons. Following this lead, in this paper we therefore study
    models that modify the $b\to s\mu^+\mu^-$ transition while leaving
    the $b\to s e^+ e^-$ decays unchanged.}.

\item The differential branching ratio of $B^0 \to K^0 \mu^+
  \mu^-$. The experimental measurements \cite{Aaij:2014pli} are given in Table
  \ref{B0KmumuBRmeas} in the Appendix.

\item $\bs \to \phi \mu^+ \mu^-$: The experimental measurements of the
  differential branching ratio and the angular observables \cite{BsphimumuLHCb2} are given
  respectively in Tables \ref{BsphimumuBRmeas} and
  \ref{Bsphimumuangmeas} in the Appendix.

\item The differential branching ratio of $B \to X_s \mu^+ \mu^-$. The
  experimental measurements \cite{Lees:2013nxa} are given in Table \ref{BXsmumuBRmeas}
  in the Appendix.

\item ${\rm BR}(\bs \to \mu^+ \mu^-) = (2.9 \pm 0.7) \times 10^{-9}$
  \cite{Aaij:2013aka,CMS:2014xfa}.

\end{enumerate}

In computing the theoretical predictions for the above observables, we
note the following:
\begin{itemize}

\item For $B\to K^* \mu^+ \mu^-$ and $\bs \to \phi \mu^+ \mu^-$, we
  use the form factors from the combined fit to lattice and light-cone
  sum rules (LCSR) calculations \cite{QCDsumrules}. These
  calculations are applicable to the full $q^2$ kinematic region. In
  LCSR calculations the full error correlation matrix is used, which
  is useful to avoid an overestimate of the uncertainties.

\item In $B \to K \mu^+ \mu^-$, we use the form factors from lattice
  QCD calculations \cite{Bailey:2015dka}, in which the main sources of
  uncertainty are from the chiral-continuum extrapolation and the
  extrapolation to low $q^2$. In order to cover the entire
  kinematically-allowed range of $q^2$, we use the model-independent
  $z$ expansion given in Ref.~\cite{Bailey:2015dka}.

\item The decay $\bs \to \phi \mu^+ \mu^-$ has special
  characteristics, namely (i) there can be (time-dependent) indirect
  CP-violating effects, and (ii) the $\bs$-$\bsbar$ width difference,
  $\Delta\Gamma_s$, is non-negligible. These must be taken into
  account in deriving the angular distribution, see
  Ref.~\cite{Descotes-Genon:2015hea}. In {\tt flavio} \cite{flavio},
  the width difference is taken into account, but all observables
  correspond to time-integrated ones (so no indirect CP violation).

\item In the calculation of the branching ratio of the inclusive decay
  $B \to X_s \mu^+ \mu^-$, the dominant perturbative contributions are
  calculated up to NNLO precision following Refs.~\cite{Asatryan:2002iy,
  Ghinculov:2003qd, Huber:2005ig, Huber:2007vv}.

\end{itemize}

The above observables are used in all fits. However, a particular
model may receive further constraints from its contributions to other
observables, such as $\bsnunubar$, $\bs$-$\bsbar$ mixing, etc. These
additional constraints will be discussed when we describe the
model-dependent fits.

\subsection{\bf Predictions}

Eq.~(\ref{angdist}) applies to $B^0 \to K^{*0} \mu^+ \mu^-$ decays.
Here the seven angular observables $S_3$, $S_4$, $S_5$, $A_{FB}$,
$S_7$, $S_8$ and $S_9$ are obtained by averaging the angular
distributions of $B$ and ${\bar B}$ decays. However, one can also
consider the difference between $B$ and ${\bar B}$ decays. This leads
to seven angular asymmetries: $A_3$, $A_4$, $A_5$, $A_6^s$, $A_7$,
$A_8$ and $A_9$ \cite{BHP,BK*mumuCPV}. For $B \to K \mu^+ \mu^-$,
there is only the partial rate asymmetry $A_{\rm CP}$.

In general, there are two categories of CP asymmetries. Suppose the
two interfering amplitudes are $A_{\rm SM} = a_1 e^{i \phi_1} e^{i
  \delta_1}$ and $A_{\rm NP} = a_2 e^{i \phi_2} e^{i \delta_2}$, where
the $a_i$ are the magnitudes, the $\phi_i$ the weak phases and the
$\delta_i$ the strong phases. Direct CP asymmetries involving rates
are proportional to $\sin (\phi_1 - \phi_2) \sin (\delta_1 -
\delta_2)$. On the other hand, CP asymmetries involving T-odd triple
products of the form ${\vec p}_i \cdot ({\vec p}_j \times {\vec p}_k)$
are proportional to $\sin (\phi_1 - \phi_2) \cos (\delta_1 -
\delta_2)$. Both types of CP asymmetry are nonzero only if the
interfering amplitudes have different weak phases, but the direct CP
asymmetry requires in addition a nonzero strong-phase difference. In
the SM, the weak phase ($= {\rm arg}(V_{tb} V_{ts}^*)$) and strong
phases are all rather small, and the NP strong phase is negligible
\cite{DatLon}. From this, we deduce that (i) large CP asymmetries are
possible only if the NP weak phase is sizeable, and (ii) triple
product CP asymmetries are most promising for seeing NP since they do
not require large strong phases.

In order to compute the predictions for the CP asymmetries, we proceed
as follows.  As noted above, we start by assuming that the NP
contributes to a particular set of $\bsmumu$ WCs. We then perform fits
to determine whether this set of WCs is consistent with all
experimental data. In the case of a model-independent fit, the data
involve only $\bsmumu$ observables; a model-dependent fit may involve
additional observables.  We determine the values of the real and
imaginary parts of the WCs that minimize the $\chi^2$. In the case of
a good fit, we then use these WCs to predict the values of the
CP-violating asymmetries $A_3$-$A_9$ in $B^0 \to K^{*0} \mu^+ \mu^-$
and $A_{\rm CP}$ in $B \to K \mu^+ \mu^-$.

In Ref.~\cite{BHP}, it was noted that $A_3$, $A_4$, $A_5$ and $A_6^s$
are direct CP asymmetries, while $A_7$, $A_8$ and $A_9$ are triple
product CP asymmetries. Furthermore, $A_7$ is very sensitive to the
phase of $C_{10}$. We therefore expect that, if NP reveals itself
through CP-violating effects in $B \to K^{(*)} \mu^+ \mu^-$, it will
most likely be in $A_7$-$A_9$, with $A_7$ being particularly
promising.

\section{Model-Independent Results}

In Refs.~\cite{Altmannshofer:2014rta,BK*mumulatestfit1}, global
analyses of the $\bsll$ anomalies were performed.  It was found that
there is a significant disagreement with the SM, possibly as large as
$4\sigma$, and that it can be explained if there is NP in $b \to s
\mu^+ \mu^-$. Ref.~\cite{BK*mumulatestfit1} offered four possible
explanations, each having roughly equal goodness-of-fits:
\bea
&{\rm (I)}& C_9^{\mu\mu}({\rm NP}) < 0 ~, \nn\\
&{\rm (II)}& C_9^{\mu\mu}({\rm NP}) = - C_{10}^{\mu\mu}({\rm NP}) < 0 ~, \nn\\
&{\rm (III)}& C_9^{\mu\mu}({\rm NP}) = - C_{9}^{\prime \mu\mu}({\rm NP}) < 0 ~, \nn\\
&{\rm (IV)}& C_9^{\mu\mu}({\rm NP}) = - C_{10}^{\mu\mu}({\rm NP}) = -C_{9}^{\prime \mu\mu}({\rm NP}) = - C_{10}^{\prime \mu\mu}({\rm NP}) < 0 ~.
\label{bsmumuWCs}
\eea
In this section we apply our method to these four scenarios. There are
several reasons for doing this. First, we want to confirm
independently that, if the NP contributes to these sets of WCs, a good
fit to the data is obtained. Note also that the above solutions were
found assuming the WCs to be real. Since we allow for complex WCs,
there may potentially be differences. Second, the main idea of the
paper is that CP-violating observables can be used to distinguish the
various NP $\bsmumu$ models. We can test this hypothesis with
scenarios I-IV. Finally, it will be useful to compare the
model-independent and model-dependent fits.

\subsection{\bf Fits}

The four scenarios are model-independent, so that the fit includes
only the $\bsmumu$ observables. The results are shown in Table
\ref{micouplins}. In scenarios II and III, there are two best-fit
solutions, labeled (A) and (B). In both cases, the two solutions have
similar best-fit values for Re(WC), but opposite signs for the
best-fit values of Im(WC). In all cases, we obtain good fits to the
data. The pulls are all $\ge 4$, indicating significant improvement
over the SM. Indeed, our results agree entirely with those of
Ref.~\cite{BK*mumulatestfit1}.

\begin{table}[htb]
\hspace{0.5in}\begin{tabular}{|c|c|c|} \hline
Scenario & [Re(WC), Im(WC)]  & pull  \\
\hline
(I) $C_9^{\mu\mu}({\rm NP}) $
         &~~[$(-1.1  \pm 0.2) $, $(0.0 \pm 0.9) $ ] & 4.2  \\
\hline
(II) $C_9^{\mu\mu}({\rm NP})  = -C_{10}^{\mu\mu}({\rm NP}) $
         & (A)~~~~~[$(-0.8\pm 0.3) $, $(1.2 \pm 0.7) $ ] & 4.2  \\
         & (B)~~~[$(-0.8  \pm 0.3) $, $(-1.2 \pm 0.8) $ ] &4.0   \\
\hline
(III) $C_9^{\mu\mu}({\rm NP})  = -C_{9}^{'{\mu\mu}}({\rm NP}) $
         &(A)~~~~~ [$(-1.0  \pm 0.2) $, $(0.3 \pm 0.6) $ ] &4.4   \\
         &(B)~~~ [$(-0.9  \pm 0.2) $, $(-0.3 \pm 0.8) $ ] & 4.4  \\
\hline
(IV) $C_9^{\mu\mu}({\rm NP})  = -C_{10}^{\mu\mu}({\rm NP})  $
         &~~~ [$(-0.6 \pm 0.2) $, $(0.1 \pm 1.2) $ ] & 4.1  \\
         $ = -C_{9}^{'{\mu\mu}}({\rm NP}) = -C_{10}^{'{\mu\mu}}({\rm NP}) $
         &~~~ &    \\
\hline
\end{tabular}
\caption{Model-independent scenarios: best-fit values of the real and
  imaginary parts of the NP WCs, as well as the pull =
  $\sqrt{\chi^2_{SM} - \chi^2_{min}}$ for the fits. For each case
  there are 104 degrees of freedom.
\label{micouplins}}
\end{table}

\subsection{\bf CP asymmetries: predictions}

\begin{figure}[!htbp]
\hspace{-0.4in}\includegraphics[height=7.0cm,width=10.0cm]{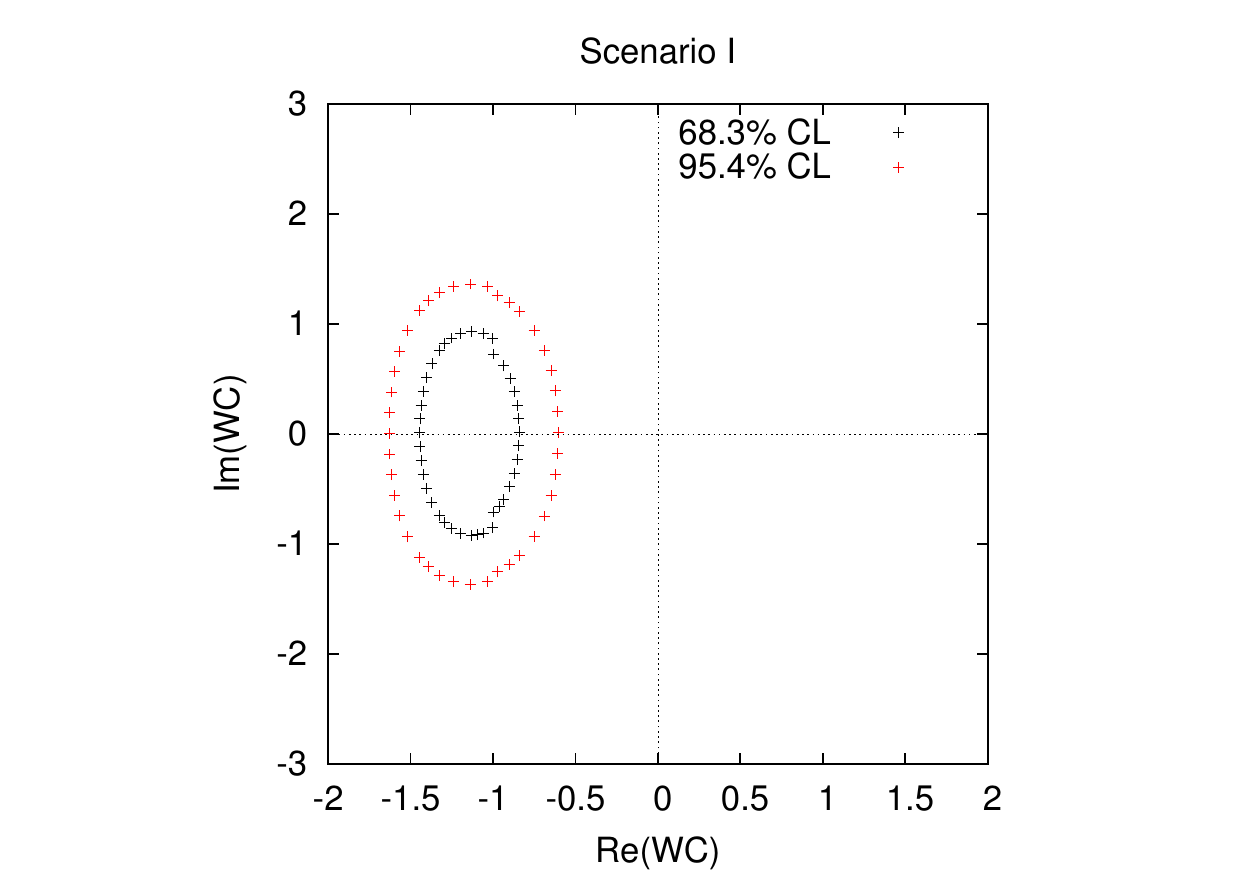}\hspace{-1.0in}
\includegraphics[height=7.0cm,width=10.0cm]{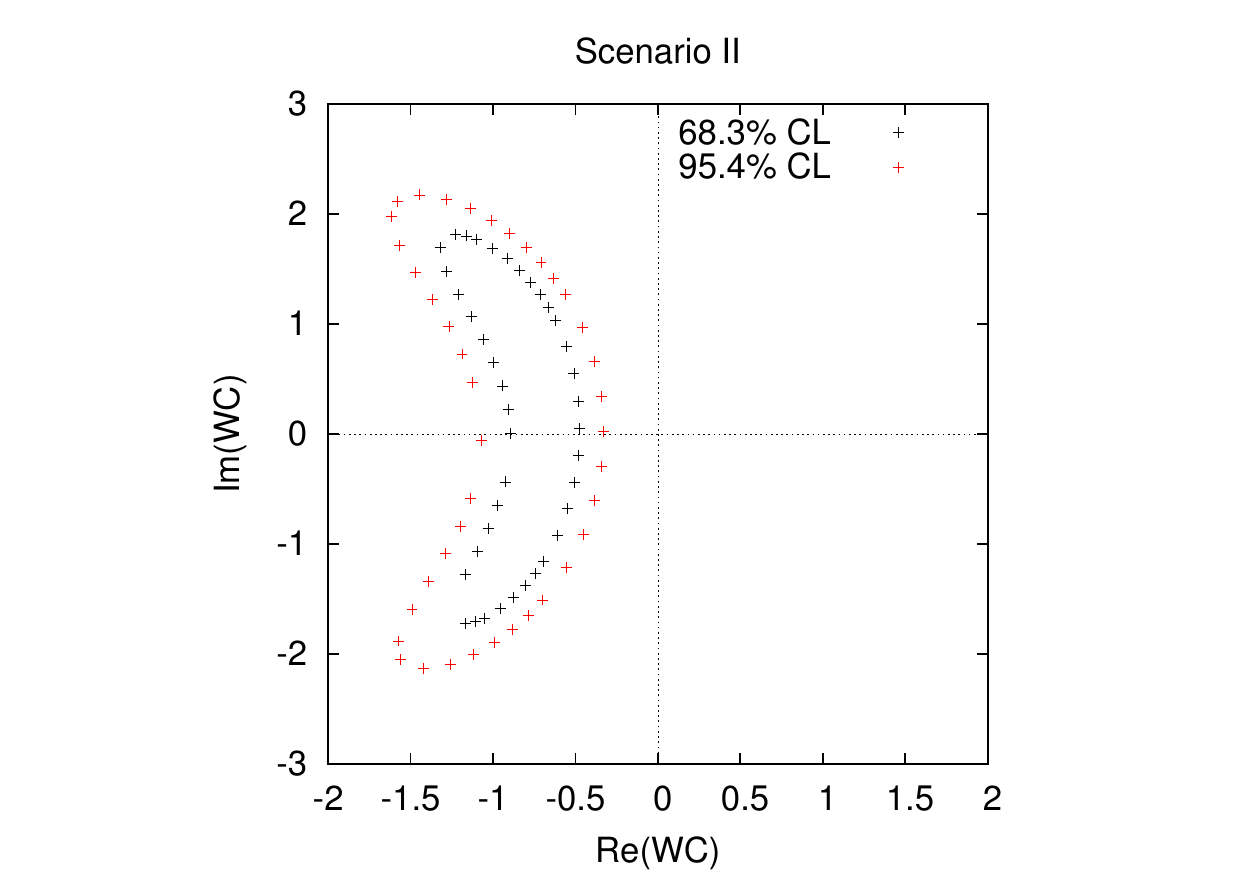}
\vspace{0.1in}
\hspace{-0.4in}\includegraphics[height=7.0cm,width=10.0cm]{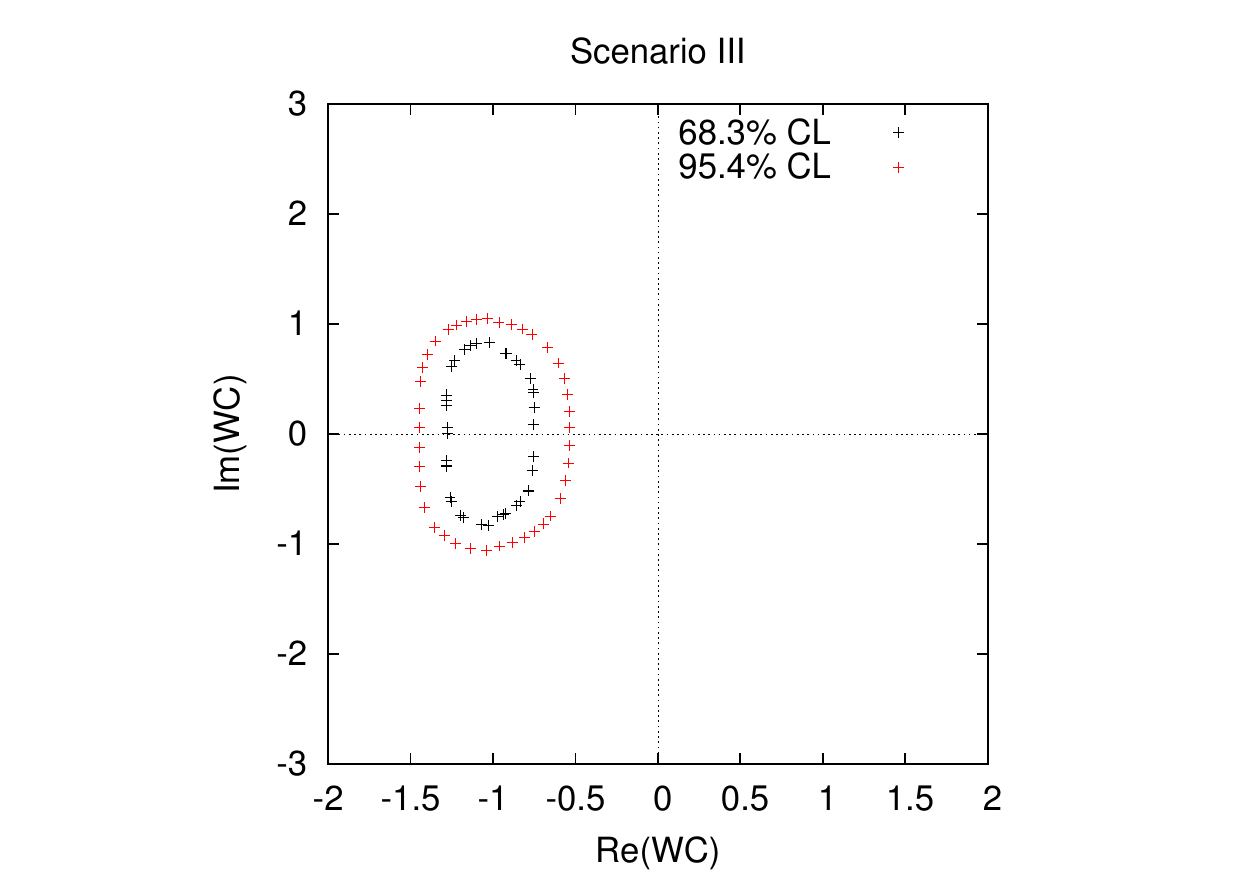}\hspace{-1.0in}
\includegraphics[height=7.0cm,width=10.0cm]{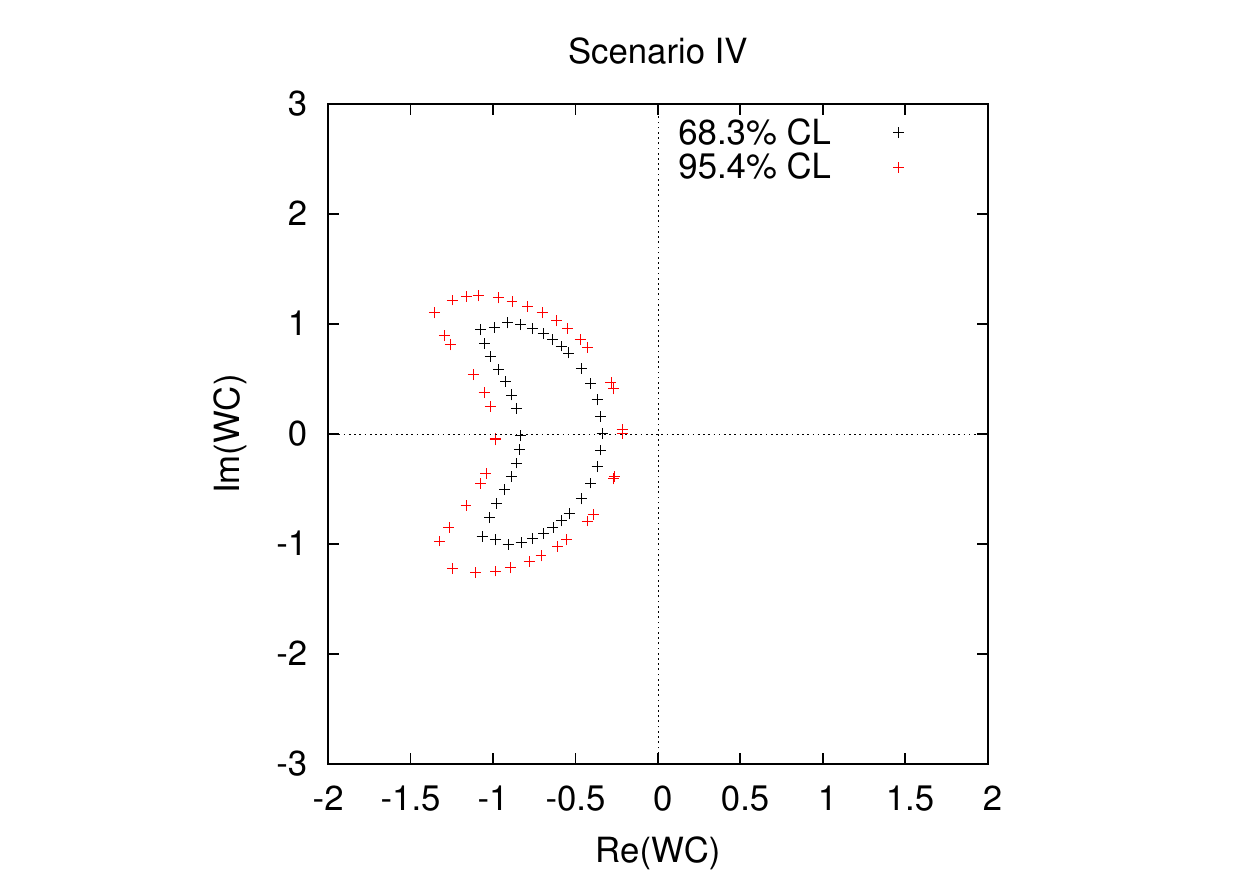}
\caption{\small Allowed regions in the Re(WC)-Im(WC) plane for the four
model-independent scenarios I-IV. See Table \ref{micouplins} for definitions of
Re(WC) and Im(WC) in each of the four scenarios.}
\label{fig:I-II}
\end{figure}

\begin{figure}[t]
\centering
\includegraphics[height=6.5cm,width=11.0cm]{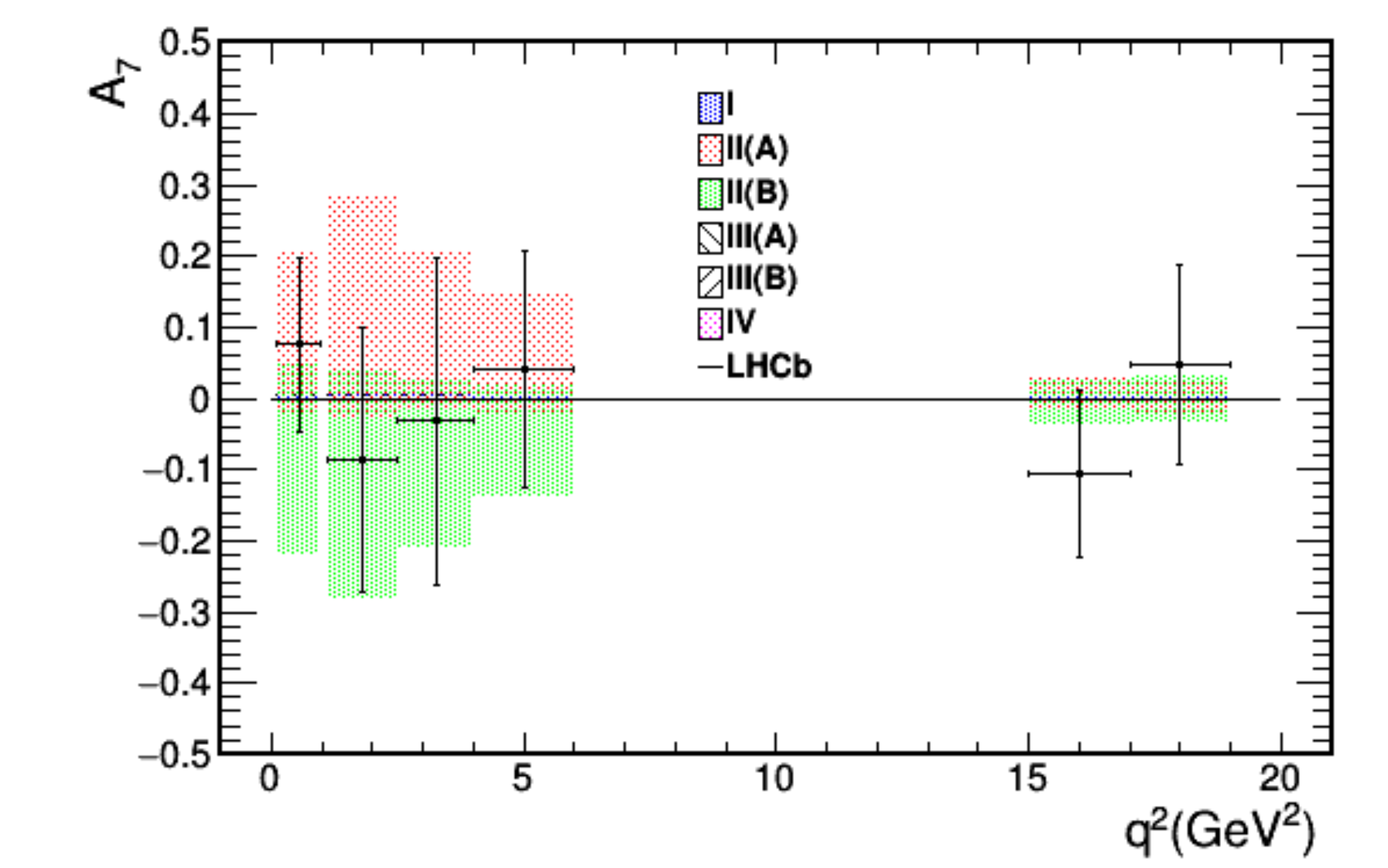}
\vspace{0.1in}
\includegraphics[height=6.5cm,width=11.0cm]{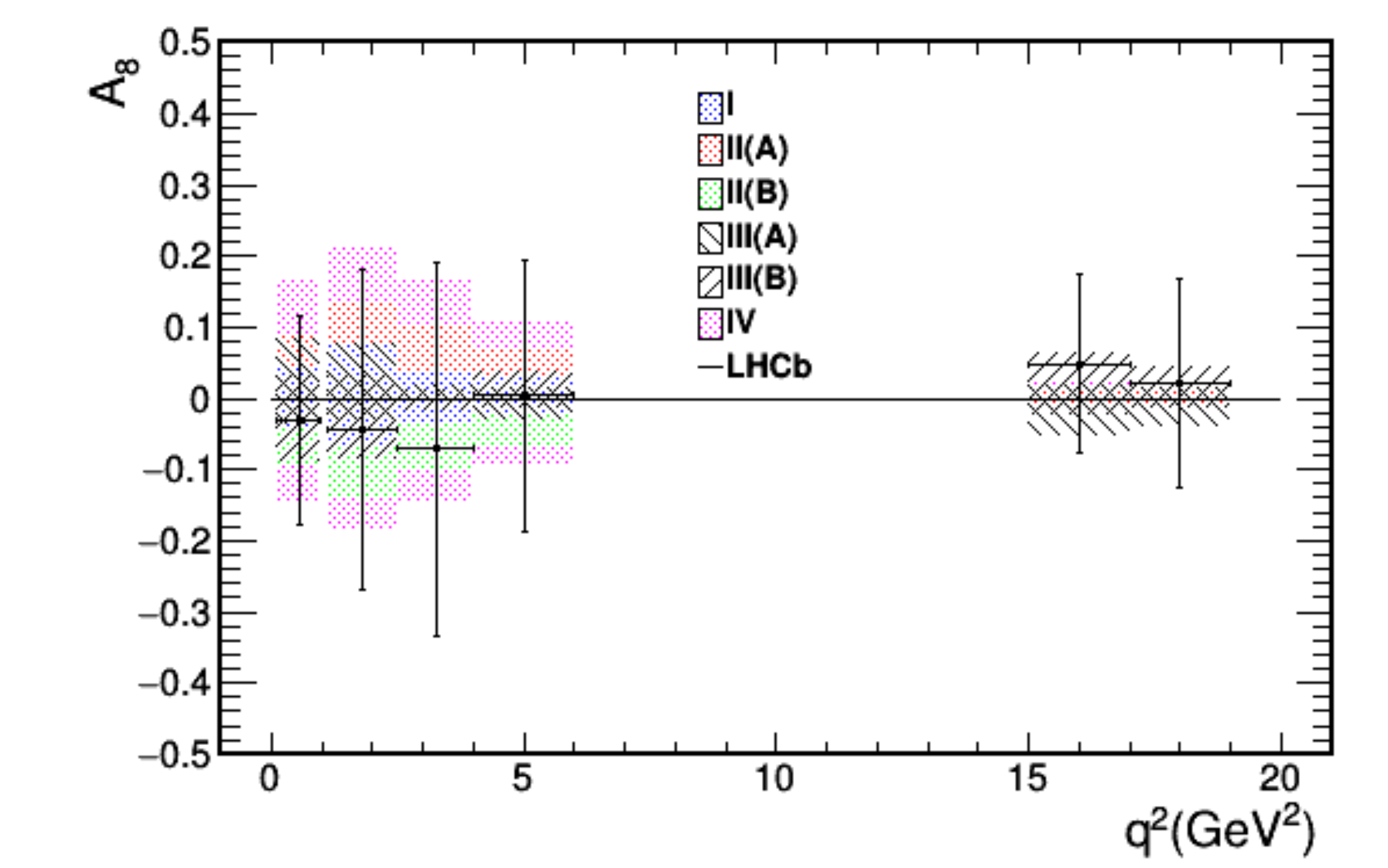}
\vspace{0.2in}
\includegraphics[height=6.5cm,width=11.0cm]{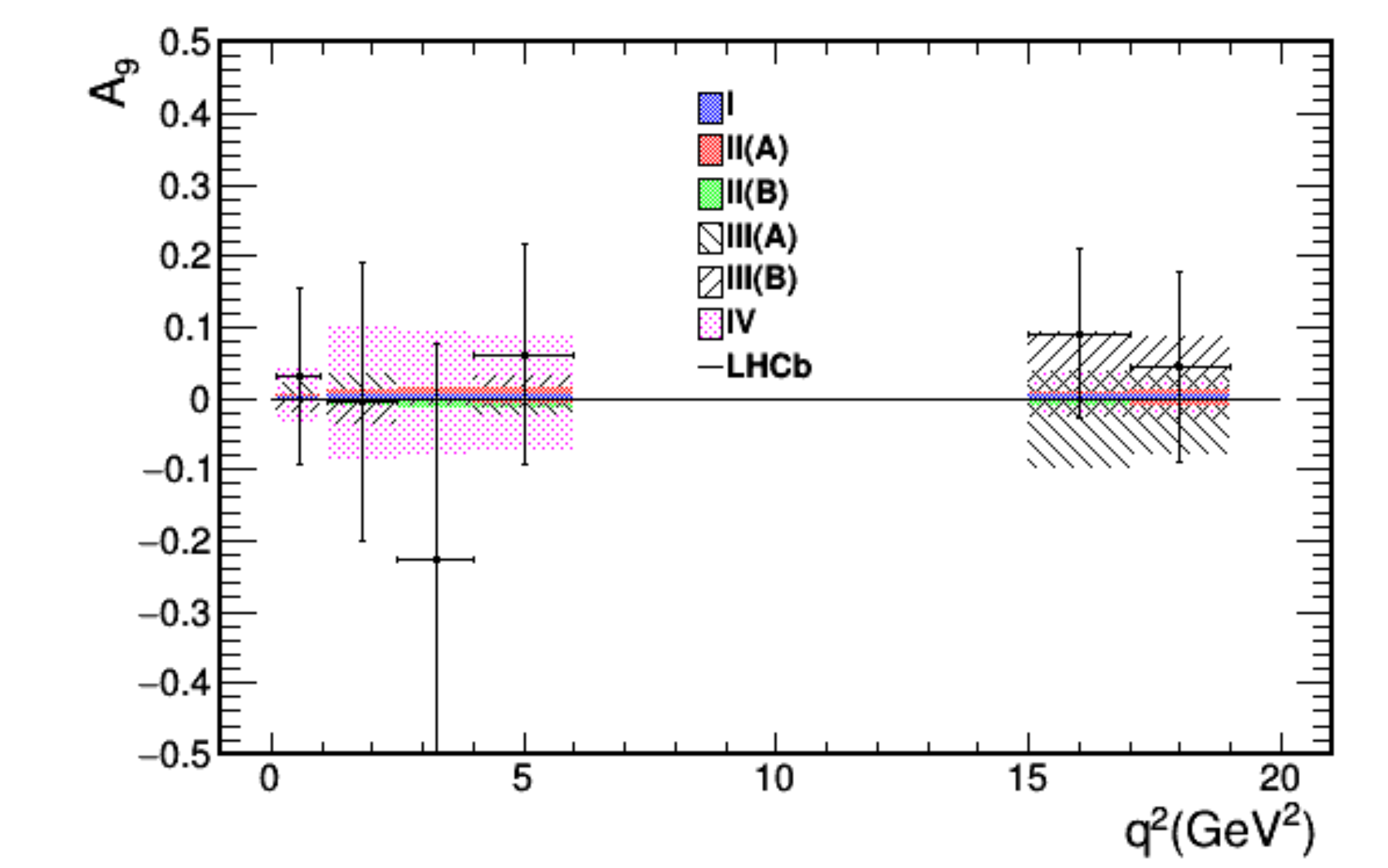}
\caption{\small Predictions of the CP asymmetries $A_7$, $A_8$ and
  $A_9$ at the $2\sigma$ level for the four model-independent scenarios
  I-IV.}
\label{fig:a9}
\end{figure}

For each of the four scenarios, the allowed values of Re(WC) and
Im(WC) are shown in Fig.~\ref{fig:I-II}. In all cases, Im(WC) is
consistent with 0, but large non-zero values are still allowed. Should
this happen, significant CP-violating asymmetries in $B \to K^{(*)}
\mu^+ \mu^-$ can be generated. To illustrate this, for each of the
four scenarios, we compute the predicted values of the CP asymmetries
$A_7$, $A_9$ and $A_8$ in $B^0 \to K^{*0} \mu^+ \mu^-$. The results
are shown in Fig.~\ref{fig:a9}. From these plots, one sees that, in
principle, one can distinguish all scenarios. If a large $A_7$
asymmetry is observed, this indicates scenario II, and one can
differentiate solutions (A) and (B). A large $A_9$ asymmetry at low
$q^2$ indicates scenario IV, while a large $A_9$ asymmetry at high
$q^2$ indicates scenario III (here solutions (A) and (B) can be
differentiated). Finally, if no $A_7$ or $A_9$ asymmetries are
observed, but a sizeable $A_8$ asymmetry is seen at low $q^2$, this
would be due to scenario I.

This then confirms the hypothesis that CP-violating observables can
potentially be used to distinguish the various NP models proposed to
explain the $\bsmumu$ anomalies. This said, one must be careful not to
read too much into the model-independent results. If NP is present in
$\bsmumu$ decays, it is due to a specific model. And this model may
have other constraints, either theoretical or experimental, that may
significantly change the predictions. That is, since the
model-independent fits have the fewest constraints, the CP-violating
effects shown in Fig.~\ref{fig:a9} are the largest possible. In a
particular model, there may be additional constraints, which will
reduce the predicted sizes of the CP asymmetries. For this reason,
while a model-independent analysis is useful to get a general idea of
what is possible, real predictions require a model-dependent analysis.
We turn to this in the following sections.

\section{Model-dependent Fits}

Many models have been proposed to explain the $\bsmumu$ anomalies, of
both the LQ \cite{CCO,AGC,HS1,GNR,VH,SM,FK,BFK,BKSZ} and $Z'$
\cite{CCO,Crivellin:2015lwa,Isidori,dark,Chiang,Virto,GGH,BG,BFG,Perimeter,CDH,SSV,CHMNPR,CMJS,BDW,FNZ,AQSS,CFL,Hou,CHV,CFV,CFGI,IGG,BdecaysDM,Bhatia:2017tgo}
variety. Rather than considering each model individually, in this
section we perform general analyses of the two types of models. The
aim is to answer two questions. First, what are the properties of
models required in order to provide good fits to the $\bsmumu$ data?
Second, which of these good-fit models can also generate sizeable
CP-violating asymmetries in $B \to K^{(*)} \mu^+ \mu^-$? We separately
examine LQ and $Z'$ models.

\subsection{\bf Leptoquarks}

The list of all possible LQ models that couple to SM particles through
dimension $\le 4$ operators can be found in Ref.~\cite{AGC}. There are
five spin-0 and five spin-1 LQs, denoted $\Delta$ and $V$
respectively, with couplings
\bea
{\cal L}_\Delta & = & ( y_{\ell u} {\bar\ell}_L u_R + y_{eq}\, {\bar e}_R i \tau_2 q_L ) \Delta_{-7/6}
+ y_{\ell d}\, {\bar\ell}_L d_R \Delta_{-1/6}
+ ( y_{\ell q}\, {\bar\ell}^c_L i \tau_2 q_L + y_{eu} \, {\bar e}^c_R u_R ) \Delta_{1/3} \nn\\
&& +~y_{ed}\, {\bar e}^c_R d_R \Delta_{4/3}
+ y'_{\ell q}\, {\bar\ell}^c_L i \tau_2 {\vec \tau} q_L \cdot {\vec \Delta}'_{1/3} + h.c. \nn\\
{\cal L}_V & = & (g_{\ell q}\, {\bar\ell}_L \gamma_\mu q_L + g_{ed}\, {\bar e}_R \gamma_\mu d_R) V^\mu_{-2/3}
+ g_{eu} \, {\bar e}_R \gamma_\mu u_R V^\mu_{-5/3}
+ g'_{\ell q}\, {\bar\ell}_L \gamma_\mu {\vec \tau} q_L \cdot {\vec V}^{\prime \mu}_{-2/3} \nn\\
&& +~(g_{\ell d}\, {\bar\ell}_L \gamma_\mu d_R^c + g_{eq}\, {\bar e}_R \gamma_\mu q^c_L) V^\mu_{-5/6}
+ + g_{\ell u} \, {\bar \ell}_L \gamma_\mu u_R^c V^\mu_{1/6} + h.c.
\label{LQlist}
\eea
In the fermion currents and in the subscripts of the couplings, $q$
and $\ell$ represent left-handed quark and lepton $SU(2)_L$ doublets,
respectively, while $u$, $d$ and $e$ represent right-handed up-type
quark, down-type quark and charged lepton $SU(2)_L$ singlets,
respectively. The LQs transform as follows under $SU(3)_c \times
SU(2)_L \times U(1)_Y$:
\bea
& \Delta_{-7/6} : ({\bar 3}, 2, -7/6) ~~,~~~~
\Delta_{-1/6} : ({\bar 3}, 2, -1/6) ~~,~~~~
\Delta_{1/3} : ({\bar 3}, 1, 1/3) ~, & \nn\\
& \Delta_{4/3} : ({\bar 3}, 1, 4/3) ~~,~~~~
{\vec \Delta}'_{1/3} : ({\bar 3}, 3, 1/3) ~, & \nn\\
& V^\mu_{-2/3} : ({\bar 3}, 1, -2/3) ~~,~~~~
V^\mu_{-5/3} : ({\bar 3}, 1, -5/3) ~~,~~~~
{\vec V}^{\prime \mu}_{-2/3} : ({\bar 3}, 3, -2/3) ~, & \nn\\
& V^\mu_{-5/6} : ({\bar 3}, 2, -5/6) ~~,~~~~
V^\mu_{1/6} : ({\bar 3}, 2, -5/3) ~.
\eea
Note that here the hypercharge is defined as $Y =  Q_{em} - I_3$.

In Eq.~(\ref{LQlist}), the LQs can couple to fermions of any
generation. To specify which particular fermions are involved, we add
superscripts to the couplings. For example, $g^{\prime \mu s}_{\ell
  q}$ is the coupling of the ${\vec V}^{\prime \mu}_{-2/3}$ LQ to a
left-handed $\mu$ (or $\nu_\mu$) and a left-handed $s$. Similarly,
$y_{e q}^{\mu b}$ is the coupling of the $\Delta_{-7/6}$ LQ to a
right-handed $\mu$ and a left-handed $b$. These couplings are relevant
for $\bsmumu$ (and possibly $\bsnunubar$). Note that the
$V^\mu_{-5/3}$ and $V^\mu_{1/6}$ LQs do not contribute to $\bsll$.

A number of these LQs, and their effects on $\bsmumu$ and other
decays, have been analyzed separately. For example, in
Ref.~\cite{Sakakietal}, it was pointed out that four LQs can
contribute to ${\bar B} \to D^{(*)+} \tau^- {\bar\nu}_\tau$. They are:
a scalar isosinglet with $Y = 1/3$, a scalar isotriplet with $Y =
1/3$, a vector isosinglet with $Y = -2/3$, and a vector isotriplet
with $Y = -2/3$. These are respectively $\Delta_{1/3}$, ${\vec
  \Delta}'_{1/3}$, $V^\mu_{-2/3}$ and ${\vec V}^{\prime \mu}_{-2/3}$.
In Ref.~\cite{Sakakietal}, they are called $S_1$, $S_3$, $U_1$ and
$U_3$, respectively, and we adopt this nomenclature below.

The $S_3$ LQ has been studied in the context of $\bsmumu$ in
Refs.~\cite{HS1,GNR,VH,SM}. $U_1$ has been examined in
Refs.~\cite{CCO,RKRDmodels}. In Ref.~\cite{FK}, the $U_3$ LQ was
proposed as an explanation of the $\bsmumu$ anomalies.  Finally, in
Refs.~\cite{BFK,BKSZ} it was claimed that the tree-level exchange of a
$\Delta_{-1/6}$ LQ can account for the $\bsmumu$ results.

There are therefore quite a few LQ models that contribute to
$\bsmumu$, several of which have been proposed as explanations of the
$B$-decay anomalies. We would like to have a definitive answer to the
following question: which of the LQs in Eq.~(\ref{LQlist}) can
actually explain the $\bsmumu$ anomalies? Rather than rely on previous
work, we perform an independent analysis ourselves.

\subsubsection{\bf LQ fits}

The difference between model-independent and model-dependent fits is
that, within a particular model, there may be contributions to new
observables and/or new operators, and this must be taken into account
in the fit. In the case of LQ models, the LQs contribute to a variety
of operators. In addition to $O^{(\prime)}_{9,10}$ [Eq.~(\ref{Heff})],
there may be contributions to
\bea
& O^{(\prime)}_\nu = [ {\bar s} \gamma_\mu P_{L(R)} b ] [ {\bar\nu}_\mu \gamma^\mu (1 - \gamma_5) \nu_\mu ] ~, & \nn\\
& O^{(\prime)}_S = [ {\bar s} P_{R(L)} b ] [ {\bar\mu} \mu ] ~~,~~~~
O^{(\prime)}_P = [ {\bar s} P_{R(L)} b ] [ {\bar\mu} \gamma_5 \mu ] & ~.
\label{newops}
\eea
$O^{(\prime)}_\nu$ contributes to $b \to s \nu_\mu {\bar \nu}_\mu$,
while $O^{(\prime)}_S$ and $O^{(\prime)}_P$ are additional contributions
to $\bsmumu$. Based on the couplings in Eq.~(\ref{LQlist}), it is
straightforward to work out which Wilson coefficients are affected by
each LQ. These are shown in Table~\ref{LQWC} \cite{AGC}. Although the
scalar LQs do not contribute to $O^{(\prime)}_{S,P}$, some vector LQs
do. For these we have $C_P^{\mu\mu}({\rm NP}) = -C_S^{\mu\mu}({\rm NP})$
and $C_P^{\prime\mu\mu}({\rm NP}) = C_S^{\prime\mu\mu}({\rm NP})$.

\begin{table}
\begin{center}
\begin{tabular}{|c|cccc|} \hline
LQ & $C_9^{\mu\mu}({\rm NP})$ & $C_{10}^{\mu\mu}({\rm NP})$ & $C_9^{\prime\mu\mu}({\rm NP})$ & $C_{10}^{\prime\mu\mu}({\rm NP})$ \\
   & $C_S^{\mu\mu}({\rm NP})$ & $C_S^{\prime\mu\mu}({\rm NP})$ & $C_\nu^{\mu\mu}({\rm NP})$ & $C_\nu^{\prime\mu\mu}({\rm NP})$ \\
\hline
$\Delta_{1/3}  ~[S_1]$ & 0 & 0 & 0 & 0 \\
& 0 & 0 & $\frac12 y_{\ell q}^{\mu b} (y_{\ell q}^{\mu s})^*$ & 0 \\
\hline
${\vec \Delta}'_{1/3} ~[S_3]$ & $y_{\ell q}^{\prime \mu b} (y_{\ell q}^{\prime \mu s})^*$
                             & $- y_{\ell q}^{\prime \mu b} (y_{\ell q}^{\prime \mu s})^*$ & 0 & 0 \\
& 0 & 0 & $\frac12 y_{\ell q}^{\prime \mu b} (y_{\ell q}^{\prime \mu s})^*$ & 0 \\
\hline
$\Delta_{-7/6}$ & $-\frac12 y_{e q}^{\mu b} (y_{e q}^{\mu s})^*$ & $-\frac12 y_{e q}^{\mu b} (y_{e q}^{\mu s})^*$ & 0 & 0 \\
& 0 & 0 & 0 & 0 \\
\hline
$\Delta_{-1/6}$ & 0 & 0 & $-\frac12 y_{\ell d}^{\mu b} (y_{\ell d}^{\mu s})^*$ & $\frac12 y_{\ell d}^{\mu b} (y_{\ell d}^{\mu s})^*$ \\
& 0 & 0 & 0 & $-\frac12 y_{\ell d}^{\mu b} (y_{\ell d}^{\mu s})^*$ \\
\hline
$\Delta_{4/3}$ & 0 & 0 & $\frac12 y_{e d}^{\mu b} (y_{e d}^{\mu s})^*$ & $\frac12 y_{e d}^{\mu b} (y_{e d}^{\mu s})^*$ \\
& 0 & 0 & 0 & 0 \\
\hline
$V^\mu_{-2/3} ~[U_1]$ & $- g_{\ell q}^{\mu b} (g_{\ell q}^{\mu s})^*$ & $ g_{\ell q}^{\mu b} (g_{\ell q}^{\mu s})^*$
& $- g_{e d}^{\mu b} (g_{e d}^{\mu s})^*$ & $- g_{e d}^{\mu b} (g_{e d}^{\mu s})^*$ \\
& $2 g_{\ell q}^{\mu b} (g_{e d}^{\mu s})^*$ & $2 (g_{\ell q}^{\mu s})^* g_{e d}^{\mu b}$ & 0 & 0 \\
\hline
${\vec V}^{\prime \mu}_{-2/3} ~[U_3]$ & $- g_{\ell q}^{\prime \mu b} (g_{\ell q}^{\prime \mu s})^*$
                                   & $ g_{\ell q}^{\prime \mu b} (g_{\ell q}^{\prime \mu s})^*$ & 0 & 0 \\
& 0 & 0 & $- 2 g_{\ell q}^{\prime \mu b} (g_{\ell q}^{\prime \mu s})^*$ & 0 \\
\hline
$V^\mu_{-5/6}$ & $g_{e q}^{\mu s} (g_{e q}^{\mu b})^*$ & $ g_{e q}^{\mu s} (g_{e q}^{\mu b})^*$
& $g_{\ell d}^{\mu s} (g_{\ell d}^{\mu b})^*$ & $- g_{\ell d}^{\mu s} (g_{\ell d}^{\mu b})^*$ \\
& $2 g_{\ell d}^{\mu s} (g_{e q}^{\mu b})^*$ & $2 (g_{\ell d}^{\mu b})^* g_{e q}^{\mu s}$ & 0 & $g_{\ell d}^{\mu s} (g_{\ell d}^{\mu b})^*$ \\
\hline
\end{tabular}
\end{center}
\caption{Contributions of the different LQs to the Wilson coefficients
  of various operators. The normalization $K \equiv \pi / (\sqrt{2}
  \alpha G_F V_{tb} V_{ts}^* M_{LQ}^2)$ has been factored out. For
  $M_{LQ} = 1$ TeV, $K = -644.4$.
\label{LQWC}}
\end{table}

There are several observations one can make from this Table. First,
not all of the LQs contribute to $\bsmumu$: $\Delta_{1/3}$ contributes
only to $\bsnunubar$. Second, $U_1$ has two couplings, $g_{\ell q}$
and $g_{e d}$. If both are allowed simultaneously, scalar operators
are generated, and these can also contribute to $\bsmumu$. This must
be taken into account in the model-dependent fits. The situation is
similar for $V^\mu_{-5/6}$. Finally, the $S_3$ and $U_3$ LQs both have
$C_9^{\mu\mu}({\rm NP}) = -C_{10}^{\mu\mu}({\rm NP})$; they are
differentiated only by their contributions to $C_\nu^{\mu\mu}({\rm
  NP})$.

At this stage, we can perform model-dependent fits to determine which
of the LQ models can explain the data. First of all, the SM alone does
not provide a good fit. We find, for 106 degrees of freedom, that
\beq
\chi^2_{SM}/d.o.f.\ = 1.34 ~~,~~~~ {\hbox{p-value $=0.01$.}}
\eeq
We therefore confirm that the $\bsmumu$ anomalies suggest the presence
of NP.

For the scalar LQs, the results of the fits using only the $\bsmumu$
data are shown in Table \ref{scalarLQs} (we address the $\bsnunubar$
data below). For the $S_3$ LQ, there are two best-fit solutions,
labeled (A) and (B). (The two solutions have the same best-fit values
for Re(coupling), but opposite signs for the best-fit values of
Im(coupling).) From this Table, we see that only the $S_3$ LQ provides
an acceptable fit to the data. Despite the claims of
Refs.~\cite{BFK,BKSZ}, the $\Delta_{-1/6}$ LQ does not explain the
$\bsmumu$ anomalies.

\begin{table}
\begin{center}
\begin{tabular}{|c|c|c|c|c|}
\hline
LQ & Coupling & [Re(coupling), Im(coupling)] $\times 10^3$ & pull \\
\hline
${\vec \Delta}'_{1/3} ~[S_3]$ & $y_{\ell q}^{\prime \mu b} (y_{\ell q}^{\prime \mu s})^*$
         & (A)~~~[$(1.5  \pm 0.5) $, $(-1.9 \pm 1.2) $ ] &4.2   \\
& & (B)~~~[$(1.4  \pm 0.5) $, $(1.7 \pm 1.3) $ ] & 4.0  \\
\hline
$\Delta_{-7/6}$ & $y_{e q}^{\mu b} (y_{e q}^{\mu s})^*$
         & [$(0.1  \pm 0.7) $, $(0.0 \pm 1.3) $ ] & 0.1  \\
\hline
$\Delta_{-1/6}$ & $y_{\ell d}^{\mu b} (y_{\ell d}^{\mu s})^*$
         & [$(-0.1 \pm 0.3) $, $(-0.1 \pm 1.3) $ ] & 0.4 \\
\hline
$\Delta_{4/3}$ & $y_{e d}^{\mu b} (y_{e d}^{\mu s})^*$

         &~~~[$(0.2 \pm 0.7) $, $(0.0 \pm 0.9) $ ] & 0.2  \\
\hline
\end{tabular}
\end{center}
\caption{Scalar LQs: best-fit values of the real and imaginary parts
  of the couplings, and the pull=$\sqrt{\chi^2_{SM} -\chi^2_{min} }$
  of the fits, for $M_{LQ} = 1$ TeV.
\label{scalarLQs}}
\end{table}

The vector LQs are more complicated because the $U_1$ and
$V^\mu_{-5/6}$ LQs each have two couplings. The $U_1$ case, where the
two couplings are $g_{\ell q}$ and $g_{e d}$, is particularly
interesting. If $g_{e d}^{ij} = 0$, we have $C_9^{\mu\mu}({\rm NP}) =
- C_{10}^{\mu\mu}({\rm NP})$, like the $S_3$ and $U_3$ LQs. (Recall
that we found that $S_3$ can explain the $\bsmumu$ anomalies.) And if
$g_{e d}^{\mu b} (g_{e d}^{\mu s})^* = - g_{\ell q}^{\mu b} (g_{\ell
  q}^{\mu s})^*$, we have $C_9^{\mu\mu}({\rm NP}) = -
C_{10}^{\mu\mu}({\rm NP}) = -C_{9}^{\prime \mu\mu}({\rm NP}) = -
C_{10}^{\prime \mu\mu}({\rm NP})$, which is scenario IV of
Eq.~(\ref{bsmumuWCs}), and is also found to explain the anomalies. To
explore the $U_1$ model fully, we perform three fits. Fit (1) has
$g_{e d}^{ij} = 0$, fit (2) has $g_{e d}^{\mu b} = g_{\ell q}^{\mu b}$
and $g_{e d}^{\mu s} = - g_{\ell q}^{\mu s}$ (which gives $g_{e
  d}^{\mu b} (g_{e d}^{\mu s})^* = - g_{\ell q}^{\mu b} (g_{\ell
  q}^{\mu s})^*$), and fit (3) allows the $g_{e d}^{ij}$ to be
free. For the $V^\mu_{-5/6}$ LQ, here too we can allow all couplings
to vary, but for simplicity we set $g_{l d}^{ij} = 0$. However, we
have checked that, even if we vary all the couplings, this model does
not provide a good fit.

Regarding fit (3), a few comments are useful. Although we allow all
couplings to vary, the constraints apply only to products of couplings.
This allows some freedom: the magnitude of $g_{\ell q}^{\mu s}$ does
not affect the best-fit values of the WCs, so we simply set it to 1.
Also, in order to avoid problems with correlations in the fits, we set
$g_{\ell q}^{\mu s}$ and $g_{ed}^{\mu s}$ to fixed real values. Finally,
in Ref.~\cite{BK*mumulatestfit1} it was found that the global fit
requires $C_S^{\mu \mu}({\rm NP}) \ll C_9^{\mu \mu}({\rm NP})$, i.e.,
$g_{ed}^{\mu s}/ g_{\ell q}^{\mu s} \ll 1$.  We have found that
$g_{ed}^{\mu s}/g_{\ell q}^{\mu s}  \simeq 0.02$ leads to a fit with a pull of
around 4.

\begin{table}
\begin{center}
\begin{tabular}{|c|c|c|c|c|}
\hline
LQ & Couplings & [Re(coupling), Im(coupling)] $\times 10^3$ & pull \\
\hline
$V^\mu_{-2/3} ~[U_1]$: & & & \\
 (1)& $g_{\ell q}^{\mu b} (g_{\ell q}^{\mu s})^*$
         &(A)~~[$(-1.5  \pm 0.5) $, $(1.9 \pm 1.2) $ ] & 4.2   \\
& & (B)~~~[$(-1.4  \pm 0.5) $, $(-1.7 \pm 1.3) $ ] & 4.0   \\
&&& \\

(2)&$g_{\ell q}^{\mu b} (g_{\ell q}^{\mu s} )^*$
         & [$(-0.01  \pm 0.02 ) $, $(0.0 \pm 0.02) $ ] & 0.5   \\

&&& \\
(3)&$g_{\ell q}^{\mu b}$
         &(A)~~~ [$(-1.2  \pm 0.4) $, $(1.7 \pm 1.1) $ ] &    \\

&$g_{e d}^{\mu b}$
         & [$(0.07  \pm 0.04) $, $(0.02 \pm 0.08) $ ] &  4.3  \\

&~~~~~~~
         &(B)~~~ [$(-1.3  \pm 0.4) $, $(-1.9 \pm 1.0) $ ] &    \\

&~~~~~~~
         &[$(0.06  \pm 0.05) $, $(-0.02 \pm 0.08) $ ] &  4.3  \\

\hline
${\vec V}^{\prime \mu}_{-2/3} ~[U_3]$ & $g_{\ell q}^{\prime \mu b} (g_{\ell q}^{\prime \mu s})^*$
         & (A)~~~[$(-1.5  \pm 0.5) $, $(1.9 \pm 1.2) $ ] & 4.2   \\
& & (B)~~~[$(-1.4  \pm 0.5) $, $(-1.7 \pm 1.3) $ ] & 4.0   \\
\hline
$V^\mu_{-5/6}$ &
$ g_{e q}^{\mu s}(g_{e q}^{\mu b})^*$   & [$(0.0  \pm 0.4) $, $(0.0 \pm 1.2) $ ] & 0.0  \\
\hline
\end{tabular}
\end{center}
\caption{Vector LQs: best-fit values of the real and imaginary parts
  of the couplings, and the pull=$\sqrt{ \chi^2_{SM} -\chi^2_{min}}$
  of the fits, for $M_{LQ} = 1$ TeV.
\label{vectorLQs}}
\end{table}

The results of the fits are shown in Table \ref{vectorLQs}. There are
several notable features:
\begin{enumerate}

\item We see that the $\bsmumu$ anomalies can be explained with the
  $U_1$ LQ [fit (1)] and the $U_3$ LQ. Like the $S_3$ LQ, they have
  $C_9^{\mu\mu}({\rm NP}) = - C_{10}^{\mu\mu}({\rm NP})$. Indeed,
  because only $\bsmumu$ data were used in the fits, the fit results
  are identical for all three LQ models.

\item A good fit is also found with the $U_1$ LQ [fit (3)].  However,
  the best-fit solution has $g_{e d}^{\mu b} \simeq 0$, so that this
  is essentially the same as the $U_1$ LQ [fit (1)].

\item The $U_1$ LQ model [fit (2)] has been constructed to satisfy
  $C_9^{\mu\mu}({\rm NP}) = - C_{10}^{\mu\mu}({\rm NP}) =
  -C_{9}^{\prime \mu\mu}({\rm NP}) = - C_{10}^{\prime \mu\mu}({\rm
    NP})$. Despite this, the model does not provide a good fit of the
  $\bsmumu$ data. The reason is that, in this model, there are also
  important contributions to the scalar operators of
  Eq.~(\ref{newops}). However, the measurement of $\bs \to
  \mu^+\mu^-$ puts strong constraints on such contributions. The
  result is that one cannot explain the anomalies in $B \to K^*
  \mu^+\mu^-$, $\bs \to \phi \mu^+ \mu^-$ and $R_K$, while
  simultaneously agreeing with the measurement of $\bs \to
  \mu^+\mu^-$. This provides an explicit example of how the
  ``model-independent,'' results of Eq.~(\ref{bsmumuWCs}) do not
  necessarily apply to particular models.

\item The $V^\mu_{-5/6}$ LQ model does not provide a good fit of the
  $\bsmumu$ data.

\end{enumerate}

We therefore see that, of all the scalar and vector LQ models, only
$S_3$, $U_1$ and $U_3$ can explain the $\bsmumu$ anomalies.
Furthermore, within the context of $\bsmumu$ processes, the models are
equivalent, since they all have $C_9^{\mu\mu}({\rm NP}) = -
C_{10}^{\mu\mu}({\rm NP})$.

Finally, recall that the aim of this analysis is to differentiate
different $\bsmumu$ NP models through measurements of CP-violating
asymmetries in $B \to K^{(*)} \mu^+ \mu^-$. As noted in the
introduction, such CP asymmetries can be sizeable only if there is a
significant NP weak phase. For the LQ model, we see from Table
\ref{vectorLQs} that the real and imaginary parts of the coupling are
of similar sizes. The NP weak phase is therefore not small, so that
large CP asymmetries can be expected.

\subsubsection{\boldmath $\bsnunubar$}

Above, we have argued that the $S_3$, $U_1$ and $U_3$ LQ models are
equivalent. However, from Table~\ref{LQWC}, note that the three LQs
contribute differently to $C_\nu^{\mu\mu}({\rm NP})$, the WC
associated with $O_\nu$, the operator responsible for $b \to s \nu_\mu
{\bar \nu}_\mu$. To be specific, the $S_3$ and $U_3$ LQs have
$C_\nu^{\mu\mu}({\rm NP}) = \frac12 C_9^{\mu\mu}({\rm NP})$ and
$C_\nu^{\mu\mu}({\rm NP}) = 2 C_9^{\mu\mu}({\rm NP})$, respectively,
while the $U_1$ LQ has $C_\nu^{\mu\mu}({\rm NP}) = 0$.  This means
that, for $S_3$ and $U_3$, constraints on $C_\nu^{\mu\mu}({\rm NP})$
translate into additional constraints on $C_9^{\mu\mu}({\rm NP})$.
This then raises the question: could these three LQ solutions be
distinguished by the $\bsnunubar$ data?

The effective Hamiltonian relevant for $\bsnunubar$ is
\cite{Buras:2014fpa}
\beq
H_{\rm eff} = - \frac{\alpha G_F}{\s \pi} V_{tb} V_{ts}^* \sum_\ell
C_L^\ell (\bar s \gamma_{\mu} P_L b) (\bar \nu_\ell \gamma^{\mu}
(1-\gamma_5)\nu_\ell) ~.
\eeq
The WC contains both the SM and NP contributions: $C_L^\ell = C_L^{\rm
  SM} + C_\nu^{\ell\ell}({\rm NP})$; it allows for NP that is lepton flavor
non-universal. This is appropriate to the present case, as the LQs
have only a nonzero $C_\nu^{\mu\mu}({\rm NP})$. The SM WC is
\beq
C_L^{\rm SM} = - X_t/s_W^2 ~,
\eeq
where $s_W \equiv \sin\theta_W$ and $X_t = 1.469 \pm 0.017 $.

The latest $\bsnunubar$ measurements yield \cite{Grygier:2017tzo}
\bea
{\cal B}(B \rightarrow K \nu \bar \nu ) &<& 1.6 \times 10^{-5} ~, \nn\\
{\cal B}(B \rightarrow K^* \nu \bar \nu ) &<& 2.7 \times 10^{-5} ~.
\label{eq:bknunuexp}
\eea
In Ref.~\cite{Buras:2014fpa}, the SM predictions for these decays were
computed:
\bea
{\cal B}(B \rightarrow K \nu \bar \nu )\vert_{SM} = (3.98 \pm 0.43 \pm 0.19) \times 10^{-6} ~, \nn\\
{\cal B}(B \rightarrow K^* \nu \bar \nu )\vert_{SM} = (9.19 \pm 0.86 \pm 0.50) \times 10^{-6} ~.
\label{eq:bksnunuth}
\eea
We define
\beq
{\cal R}_K \equiv \frac{{\cal B}(B \rightarrow K \nu \bar \nu )}{{\cal B}_{SM}{(B \rightarrow K \nu \bar \nu)}} ~~,~~~~
{\cal R}_{K^*} \equiv \frac{{\cal B}(B \rightarrow K^* \nu \bar \nu )}{{\cal B}_{SM}{(B \rightarrow K^* \nu \bar \nu)}} ~.
\eeq
Using Eqs.~(\ref{eq:bknunuexp}) and (\ref{eq:bksnunuth}), we obtain
\beq
{\cal R}_K < 4.0 ~~,~~~~ {\cal R}_{K^*} < 2.9 ~.
\label{RKbounds}
\eeq

From Ref.~\cite{Buras:2014fpa}, ${\cal R}_K$ and ${\cal R}_{K^*}$ can
be written as
\bea
{\cal R}_K = {\cal R}_K^* & = & \frac23 + \frac13 \frac{|C_L^{SM} + C_\nu^{\mu\mu}({\rm NP})|^2}{|C_L^{SM}|^2} \nn\\
& = & 1 + \frac23 {\rm Re}(C_\nu^{\mu\mu}({\rm NP}) / C_L^{SM}) + \frac13 |C_\nu^{\mu\mu}({\rm NP}) / C_L^{SM}|^2 ~.
\label{RKWCs}
\eea
Since $C_\nu^{\mu\mu}({\rm NP})$ is proportional to $C_9^{\mu\mu}({\rm NP})$, and
since $|C_9^{\mu\mu}({\rm NP})| = O(1)$ (see Table \ref{micouplins},
scenario II), the $\bsmumu$ data implies that $|C_\nu^{\mu\mu}({\rm NP})|$
is also $O(1)$. Can the $\bsnunubar$ data provide competitive
constraints on $|C_\nu^{\mu\mu}({\rm NP})|$? Using the ${\cal R}_{K^*}$
bound of Eq.~(\ref{RKbounds}) (since it is stronger), and neglecting
${\rm Im}(C_\nu^{\mu\mu}({\rm NP}))$ in Eq.~(\ref{RKWCs}), we obtain
\beq
-10.1 < {\rm Re}(C_\nu^{\mu\mu}({\rm NP})) < 22.8 ~.
\eeq
The above limit is significantly weaker than the result
$|C_\nu^{\mu\mu}({\rm NP})| = O(1)$ coming from the fit to the $\bsmumu$
data. We therefore conclude that the $\bsnunubar$ data cannot be used
to distinguish the $S_3$, $U_1$ and $U_3$ LQs.

Note that this conclusion may not hold if the LQs also couple to other
leptons. For example, in Ref.~\cite{RKRDmodels} it was assumed that
the LQs couple to $(\nu_\tau, \tau^-)_L$ in the gauge basis, and that
couplings to $(\nu_\mu, \mu^-)_L$ are generated only when one
transforms to the mass basis. In this case, the LQs contribute not
only to $b \to s \nu_\mu {\bar \nu}_\mu$, but also to $b \to s
\nu_\tau {\bar \nu}_\tau$, which can alter the above analysis. Indeed,
in Ref.~\cite{RKRDmodels} it is found that constraints from
$\bsnunubar$ {\it are} important in the comparison of the $S_3$, $U_1$
and $U_3$ LQs.

\subsection{\bf \boldmath $Z'$ bosons}
\label{Z'bosons}

Perhaps the most obvious candidate for a NP contribution to $\bsmumu$
is the tree-level exchange of a $Z'$ boson with a flavor-changing
coupling ${\bar s} \gamma^\mu P_L b Z'_\mu$. Given that it couples to
two left-handed doublets, the $Z'$ must transform as a singlet or
triplet of $SU(2)_L$. The triplet option has been examined in
Refs.~\cite{CCO,Crivellin:2015lwa,Isidori,dark,Chiang,Virto}. (In this
case, there is also a $W'$ that can contribute to ${\bar B} \to
D^{(*)+} \tau^- {\bar\nu}_\tau$ \cite{RKRD}, another decay whose
measurement exhibits a discrepancy with the SM
\cite{RD_BaBar,RD_Belle,RD_LHCb}.) If the $Z'$ is a singlet of
$SU(2)_L$, it must be the gauge boson associated with an extra
$U(1)'$. Numerous models of this type have been proposed, see
Refs.~\cite{GGH,BG,BFG,Perimeter,CDH,SSV,CHMNPR,CMJS,BDW,FNZ,AQSS,CFL,Hou,CHV,CFV,CFGI,IGG,BdecaysDM,Bhatia:2017tgo}.

The vast majority of these $Z'$ models use scenario II of
Eq.~(\ref{bsmumuWCs}): $C_9^{\mu\mu}({\rm NP}) = -
C_{10}^{\mu\mu}({\rm NP})$. Thus, although the underlying details of
these models are different, in all cases we can write
\bea
\Delta {\cal L}_{Z'} & = & J^\mu Z'_\mu ~, \nn\\
{\rm where} \qquad J^\mu & = &
g_L^{\mu \mu} \, \bar L \gamma^{\mu} P_L L + g_L^{bs} \,
{\bar\psi}_{q2} \gamma^{\mu} P_L \psi_{q3} + h.c.
\label{Z'couplings}
\eea
Here $\psi_{qi}$ is the quark doublet of the $i^{th}$ generation, and
$L=(\nu_\mu , \mu)^T$. When the heavy $Z'$ is integrated out, we
obtain the following effective Lagrangian containing 4-fermion
operators:
\bea
{\cal L}_{Z'}^{eff} = -\frac{1}{2 M_{Z'}^2} J_\mu J^\mu
& \supset & -\frac{g_L^{bs} g_L^{\mu \mu}}{M_{Z'}^2} (\bar s \gamma^{\mu} P_L b) (\bar \mu \gamma^{\mu} P_L \mu)
- \frac{(g_L^{bs})^2}{2 M_{Z'}^2} (\bar s \gamma^{\mu} P_L b) (\bar s \gamma^{\mu} P_L b)
\nn\\
&& \hskip0.5truecm
-~\frac{(g_L^{\mu \mu})^2}{M_{Z'}^2} (\bar \mu \gamma^{\mu} P_L \mu) ({\bar \nu}_\mu \gamma^{\mu} P_L \nu_\mu) ~.
\label{Z'4fermi}
\eea
The first 4-fermion operator is relevant for $\bsmumu$ transitions,
the second operator contributes to $\bs$-$\bsbar$ mixing, and the
third operator contributes to neutrino trident production.

Note that $g_L^{\mu \mu}$ must be real, since the leptonic current of
Eq.~(\ref{Z'couplings}) is self-conjugate. However, $g_L^{bs}$ can be
complex, i.e., it can contain a weak phase. This phase can potentially
lead to CP-violating effects in $B \to K^{(*)} \mu^+ \mu^-$ via the
first 4-fermion operators of Eq.~(\ref{Z'4fermi}). The question is:
how large can this NP weak phase be? This is the question that is
addressed in this subsection by considering constraints from
$\bsmumu$, $\bs$-$\bsbar$ mixing, and neutrino trident production.

For $\bsmumu$ we have
\beq
C_9^{\mu \mu}({\rm NP}) = -C_{10}^{\mu \mu}({\rm NP}) = -\left[ \frac{\pi}{\sqrt 2 G_F \alpha V_{tb} V_{ts}^*} \right ] \,
\frac{g_L^{bs} g_L^{\mu \mu}}{M_{Z'}^2} ~.
\eeq

Turning to $\bs$-$\bsbar$ mixing, the SM contribution arises due to a
box diagram, and is given by
\beq
N C_{VLL}^{\rm SM} \, ({\bar s}_L \gamma^\mu b_L)\,({\bar s}_L \gamma_\mu b_L) ~,
\eeq
where
\beq
N = \frac{G_F^2 m_W^2}{16\pi^2} (V_{tb} V_{ts}^*)^2 ~,~~
C_{VLL}^{\rm SM} = \eta_{B_s} x_t \left[ 1 + \frac{9}{1-x_t} - \frac{6}{(1-x_t)^2} -\frac{6 x_t^2 \ln x_t}{(1-x_t)^3} \right] ~.
\eeq
Here $x_t \equiv m_t^2/m_W^2$ and $\eta_{B_s} = 0.551$ is the QCD
correction \cite{Buchalla:1995vs}.  Combining the SM and NP
contributions, we define
\beq
N C_{VLL} \equiv |N C_{VLL}^{\rm SM}| e^{-2 i \beta_s}  + \frac{(g_L^{bs})^2}{2 M_{Z'}^2} ~,
\eeq
where $-\beta_s = {\rm arg}(V_{tb} V_{ts}^*)$. This leads to
\begin{equation}
\Delta M_s =\frac{2}{3} m_{B_s} f_{B_s}^2 \hat B_{B_s}  \left | N C_{VLL} \right  | ~.
\end{equation}
In addition, the weak phase of $\bs$-$\bsbar$ mixing is given by
\begin{equation}
\varphi_s = {\rm arg}(N C_{VLL}).
\end{equation}

From the above expressions, we see that, the larger $g_L^{bs}$ is, the
more $Z'$ models contribute to -- and receive constraints from --
$\bs$-$\bsbar$ mixing. The experimental measurements of the mixing
parameters yield \cite{HFAG}
\bea
\Delta M_s^{\rm exp} & = & 17.757 \pm 0.021 ~{\rm ps}^{-1} ~, \nn\\
\varphi_s^{c{\bar c}s} &=& -0.030 \pm 0.033 ~.
\eea
These are to be compared with the SM predictions:
\bea
\Delta M_s^{\rm SM} &=& \frac{2}{3} m_{B_s} f_{B_s}^2 \hat B_{B_s}
|N C_{VLL}^{\rm SM} | = (17.9 \pm 2.4)~{\rm ps}^{-1} ~, \nn\\
\varphi_s^{c{\bar cs,{\rm SM}}} &=& -2 \beta_s = -0.03704 \pm 0.00064 ~.
\eea
In the above, for $\Delta M_s^{\rm SM}$, we have followed the
computation of Ref.~\cite{RKRDmodels}, using $ f_{B_s} \sqrt{\hat
  B_{B_s}}=270\pm 16$ MeV \cite{Aoki:2014nga,Gamiz:2009ku,Aoki:2016frl}, $|V^{}_{tb}V^*_{ts}| =
0.0405 \pm 0.0012$ \cite{pdg}, and $\overline{m_t} = 160$ GeV;
$\varphi_s^{c{\bar c}s,{\rm SM}}$ is taken from
Refs.~\cite{Charles:2004jd,Hocker:2001xe}.

The $Z'$ will also contribute to the production of $\mu^+\mu^-$ pairs
in neutrino-nucleus scattering, $\nu_\mu N \to \nu_\mu N \mu^+ \mu^-$
(neutrino trident production). At leading order, this process is
effectively $\nu_\mu \gamma \to \nu_\mu \mu^+ \mu^-$, and is produced
by single-$W$/$Z$ exchange in the SM. This arises from the
four-fermion effective operator
\beq
\mathcal L_\text{eff:trident}
 = \left[ \bar\mu \gamma^\mu \left( C_V - C_A \gamma^5 \right) \mu \right] \left[ \bar\nu \gamma_\mu (1-\gamma^5) \nu \right]\,,
\label{EQ:effectiveop}
\eeq
with an external photon coupling to $\mu^+$ or $\mu^-$. In the SM,
combining both $W$- and $Z$-exchange diagrams, we
have~\cite{Koike:1971tu,Koike:1971vg,Belusevic:1987cw,Brown:1973ih}
\beq
 C_V^\text{SM} = - {g^2 \over 8 m_W^2} \left( {1 \over 2} + 2 s_W^2 \right) ~~,~~~~
 C_A^\text{SM} = - {g^2 \over 8 m_W^2} \, {1 \over 2} ~.
\eeq
On the other hand, the $Z'$ boson contributes to
Eq.~(\ref{EQ:effectiveop}) with the pure $V-A$ form:
\beq
 C_V^{\rm NP} = C_A^{\rm NP}  = -{(g_L^{\mu \mu})^2 \over 4 M_{Z'}^2} \,.
\eeq

The theoretical prediction is then
\bea
 \left. { \sigma_\text{SM+NP} \over \sigma_\text{SM} } \right|_{\nu N \to \nu N \mu^+ \mu^-}
& = &
 { (C_V^\text{SM} + C_V^\text{NP})^2 + (C_A^\text{SM} + C_A^\text{NP})^2 \over (C_V^\text{SM})^2 + (C_A^\text{SM})^2 } \\
& = &
\frac{1}{1+(1+4s_W^2)^2} \left [ \left ( 1 + \frac{v^2(g_L^{\mu \mu})^2}{M_{{Z}^{'}}^2}  \right )^2
+ \left ( 1 +4 s_W^2 +  \frac{v^2 (g_L^{\mu \mu})^2}{M_{{Z}^{'}}^2 }  \right )^2 \right ] ~, \nn
\eea
to be compared with the experimental measurement \cite{CCFR}:
\beq
 \left. { \sigma_\text{exp.} \over \sigma_\text{SM} } \right|_{\nu N \to \nu N \mu^+ \mu^-} = 0.82 \pm 0.28 ~.
\eeq
The net effect is that this will provide an upper limit on $(g_L^{\mu
  \mu})^2/M_{Z'}^2$. For $M_{Z'}=1 $TeV and $v = 246$ GeV, we obtain the following $1\sigma$
bound on the coupling:
\begin{equation}
|g_L^{\mu\mu}| \le 1.25 ~.
\end{equation}

We now perform a fit within the context of this $Z'$ model. The fit
includes the measurements of the $\bsmumu$ observables, $\bs$-$\bsbar$
mixing (magnitude and phase), and the cross section for neutrino
trident production. There are 107 degrees of freedom.

\begin{table}[htb]
\begin{tabular}{|c|c|c|} \hline
$g_L^{\mu \mu}$ & [Re($g_L^{bs}$),Im($g_L^{bs}$)]$\times 10^{3}$  & pull  \\
\hline
0.01
         &[$(-2.4  \pm 2.1) $, $(-0.1 \pm 0.7) $ ] & 0.8   \\
         \hline
0.05
         &[$(-3.9  \pm 1.2) $, $(0.0 \pm 0.5) $ ] & 2.3   \\
         \hline
0.1
         &[$(-4.3  \pm 1.0) $, $(0.0 \pm 0.4) $ ] &3.3   \\
         \hline
0.2
         &[$(-3.9  \pm 0.8) $, $(0.0 \pm 0.5) $ ] &4.0   \\
         \hline
0.4
         &[$(-2.1  \pm 0.5) $, $(-0.1 \pm 0.8) $ ] &4.2   \\
         \hline
0.5
         & [$(-1.8  \pm 0.5) $, $(-0.1 \pm 0.9) $ ] & 4.0   \\ \hline
0.8
         &[$(-1.1 \pm 0.3) $, $(-0.1 \pm 1.5) $ ] & 4.0  \\
\hline
1.0
         & [$(-0.8  \pm 0.3) $, $(-0.4 \pm 3.1) $ ] &4.0   \\

\hline

\hline
\end{tabular}
\caption{$Z'$ model: best-fit values of the real and imaginary parts
  of $g_L^{bs}$, and the pull=$\sqrt{\chi^2_{SM} -\chi^2_{min} }$ of
  the fits, for various values of $g_L^{\mu \mu}$ and $M_{Z'} = 1$
  TeV.
\label{tab:zprime}}
\end{table}

Our results are summarized in Table \ref{tab:zprime}. We see that a
good fit is obtained for $g_L^{\mu \mu} \ge 0.1$. (Smaller values of
$g_L^{\mu \mu}$ imply larger values for $g_L^{bs}$, which are
disfavored by measurements of $\bs$-$\bsbar$ mixing.)

Once again, recall that the ultimate aim of this study is to compare
the predictions of different models for the CP-violating asymmetries
in $B \to K^{(*)} \mu^+ \mu^-$. Such asymmetries can be sizeable only
if the NP weak phase is large. However, from Table \ref{tab:zprime},
we see that Im($g_L^{bs}$)/Re($g_L^{bs}$) is O(1) only for $g_L^{\mu
  \mu} = 0.8$, 1.0. It is intermediate for $g_L^{\mu \mu} = 0.4$, 0.5,
and is small for $g_L^{\mu \mu} = 0.1$, 0.2. We therefore expect that
models with different values of $g_L^{\mu \mu}$ will predict different
values of the CP asymmetries, potentially allowing them to be
differentiated.

From the above, we see that a large NP weak phase can only be produced
in $Z'$ models if $g_L^{\mu \mu}$ is large. However, note that, while
this is a necessary condition, it is not sufficient. In a particular
$Z'$ model, it is necessary to have a mechanism whereby $g_L^{bs}$ can
have a weak phase. This is not the case for all models. As an example,
in some models, the $Z'$ couples only to ${\bar b}b$ in the gauge
basis. Its coupling constant is therefore real. The flavor-changing
coupling to ${\bar s}b$ is only generated when transforming to the
mass basis.  However, in Refs.~\cite{CCO,RKRDmodels}, this
transformation involves only the second and third generations. In
other words, it is essentially a $2 \times 2$ rotation, which is
real. In these models a weak phase in $g_L^{bs}$ cannot be generated.

\section{CP Asymmetries: Model-dependent Predictions}

In the previous section, we have identified the characteristics of NP
models that can explain the $\bsmumu$ anomalies. We have found that
there are three LQ models -- $S_3$, $U_1$, $U_3$ -- that can do this.
All have $C_9^{\mu\mu}({\rm NP}) = - C_{10}^{\mu\mu}({\rm NP})$ and so
are equivalent, as far as $\bsmumu$ processes are concerned. There is
a whole spectrum of $Z'$ models that can explain the $\bsmumu$ data.
What is required is that the $Z'$ have couplings $g_L^{bs} \, {\bar s}
\gamma^{\mu} P_L b Z'_\mu$ and $g_L^{\mu \mu} \, {\bar\mu}
\gamma^{\mu} P_L \mu Z'_\mu$, and that $g_L^{\mu \mu}$ be $\ge 0.1$.

The purpose of this paper is to investigate whether these models can
be distinguished by measurements of CP-violating asymmetries in
$\BKstarmumu$ and $\BKmumu$. To this end, the next step is then to
compute the predictions of all models for the allowed ranges of the
various asymmetries.  For the LQ and $Z'$ models, the best-fit values
and errors of the real and imaginary parts of the NP couplings are
given in Tables \ref{scalarLQs} and \ref{tab:zprime}, respectively.
(For the LQ model, the allowed region in the Re(WC)-Im(WC) plane is
shown in the upper right plot of Fig.~\ref{fig:I-II} (scenario II).)
With these we can calculate the predictions for the asymmetries for
all models.

\begin{figure}[!htbp]
\centering
{\hspace{-0.7cm}\includegraphics[height=5.3cm,width=7.95cm]{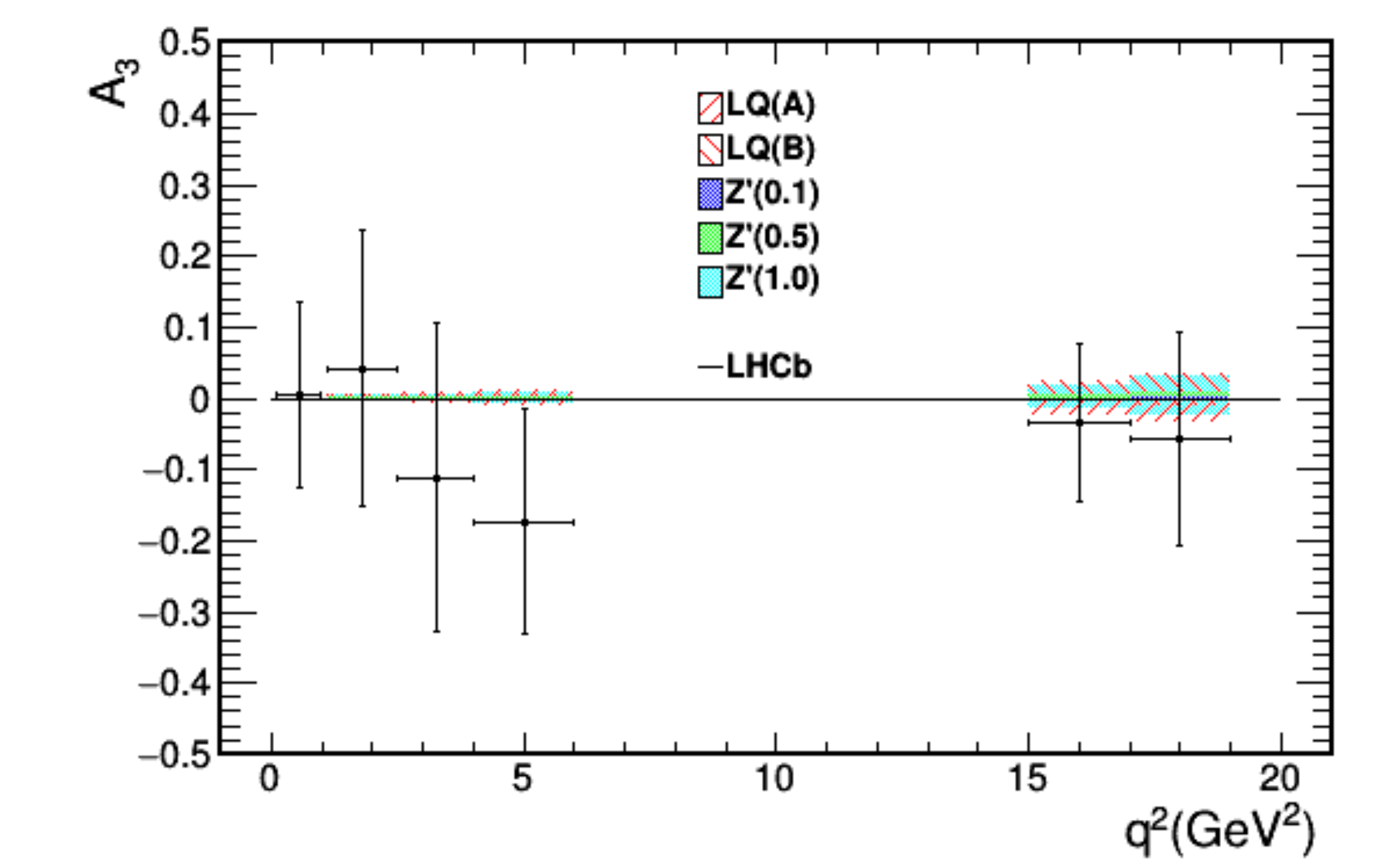}\hspace{-0.10in}
\includegraphics[height=5.3cm,width=7.95cm]{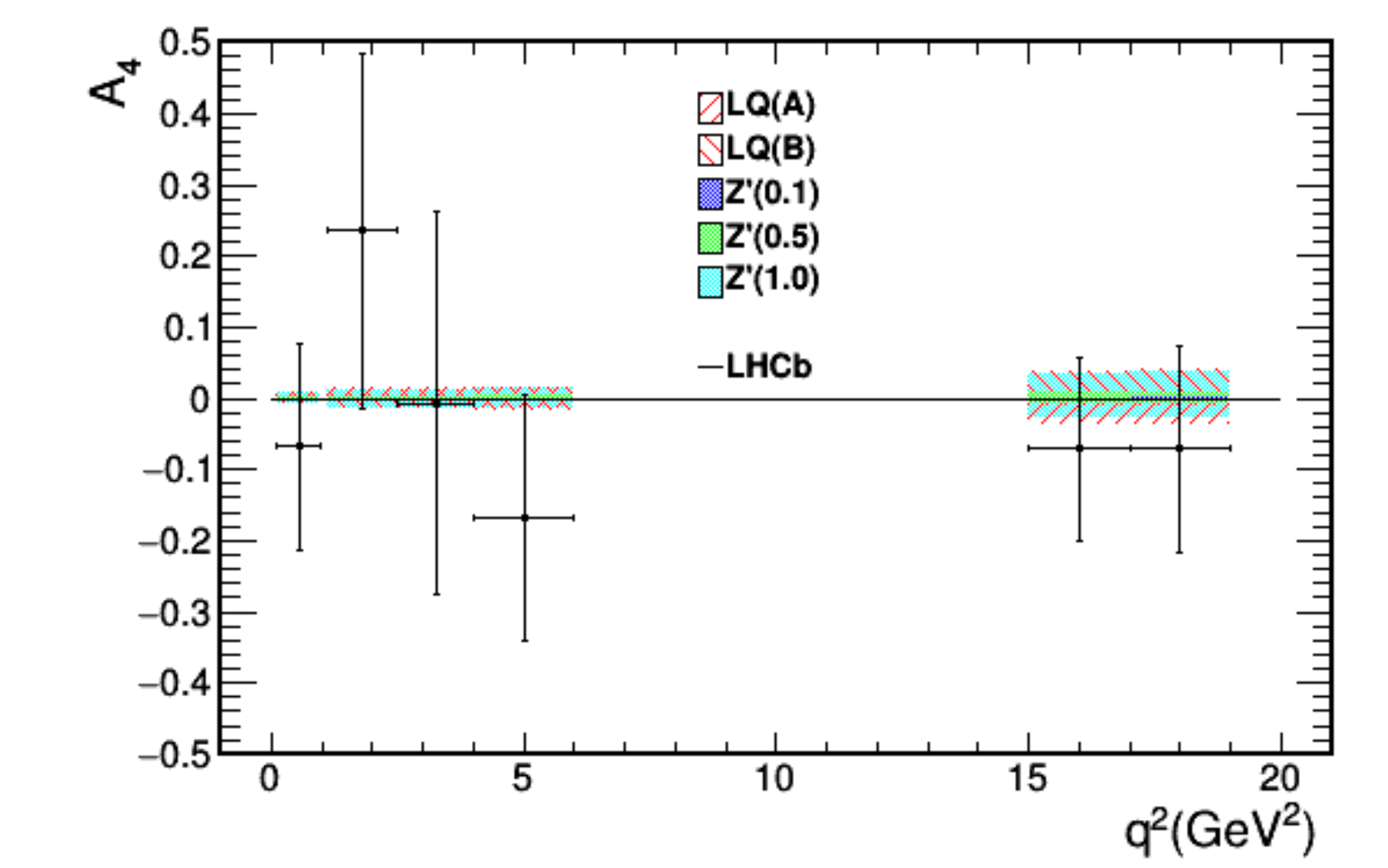}}
\vskip1truemm
{\hspace{-0.7cm}\includegraphics[height=5.3cm,width=7.95cm]{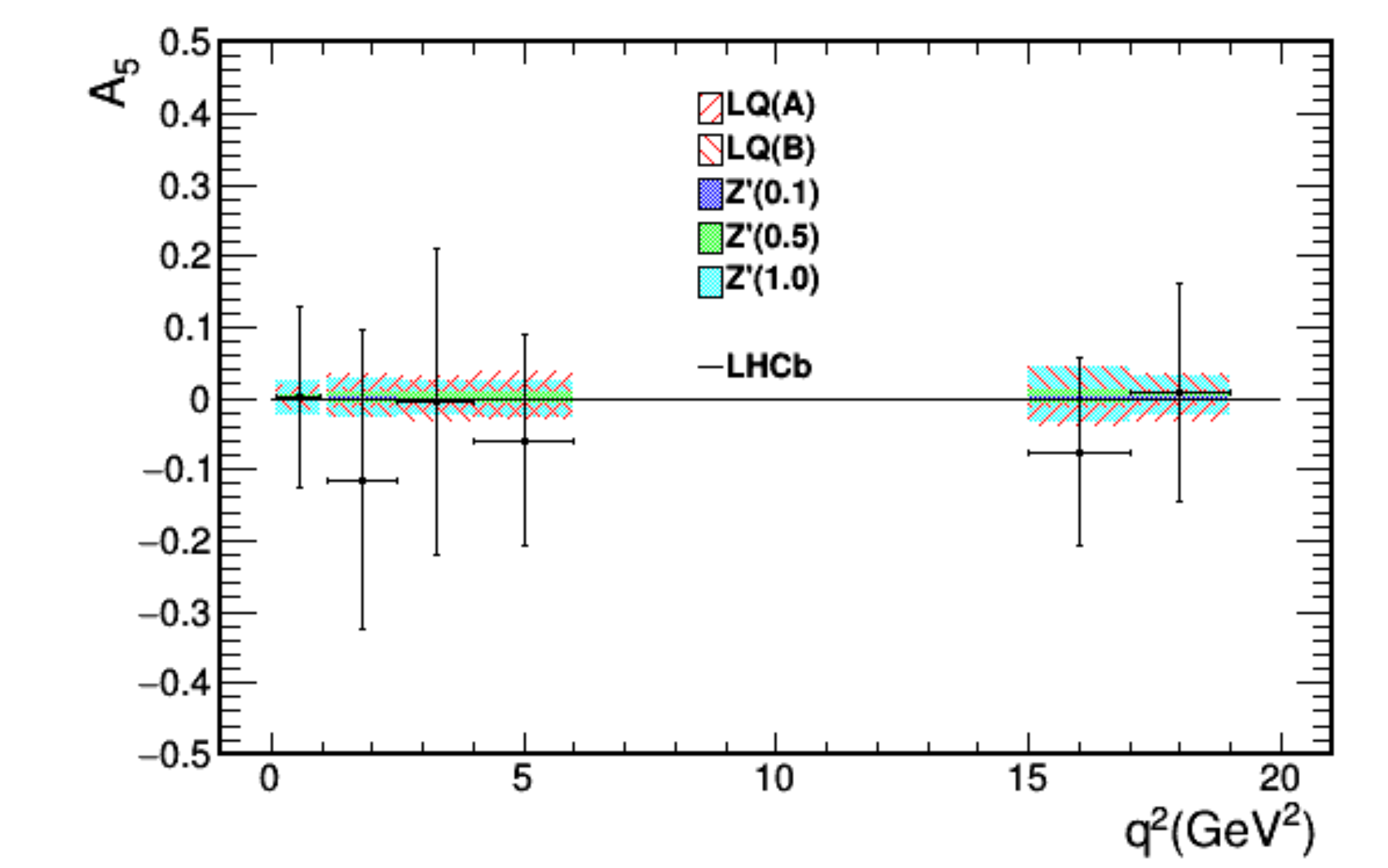}\hspace{-0.10in}
\includegraphics[height=5.3cm,width=7.95cm]{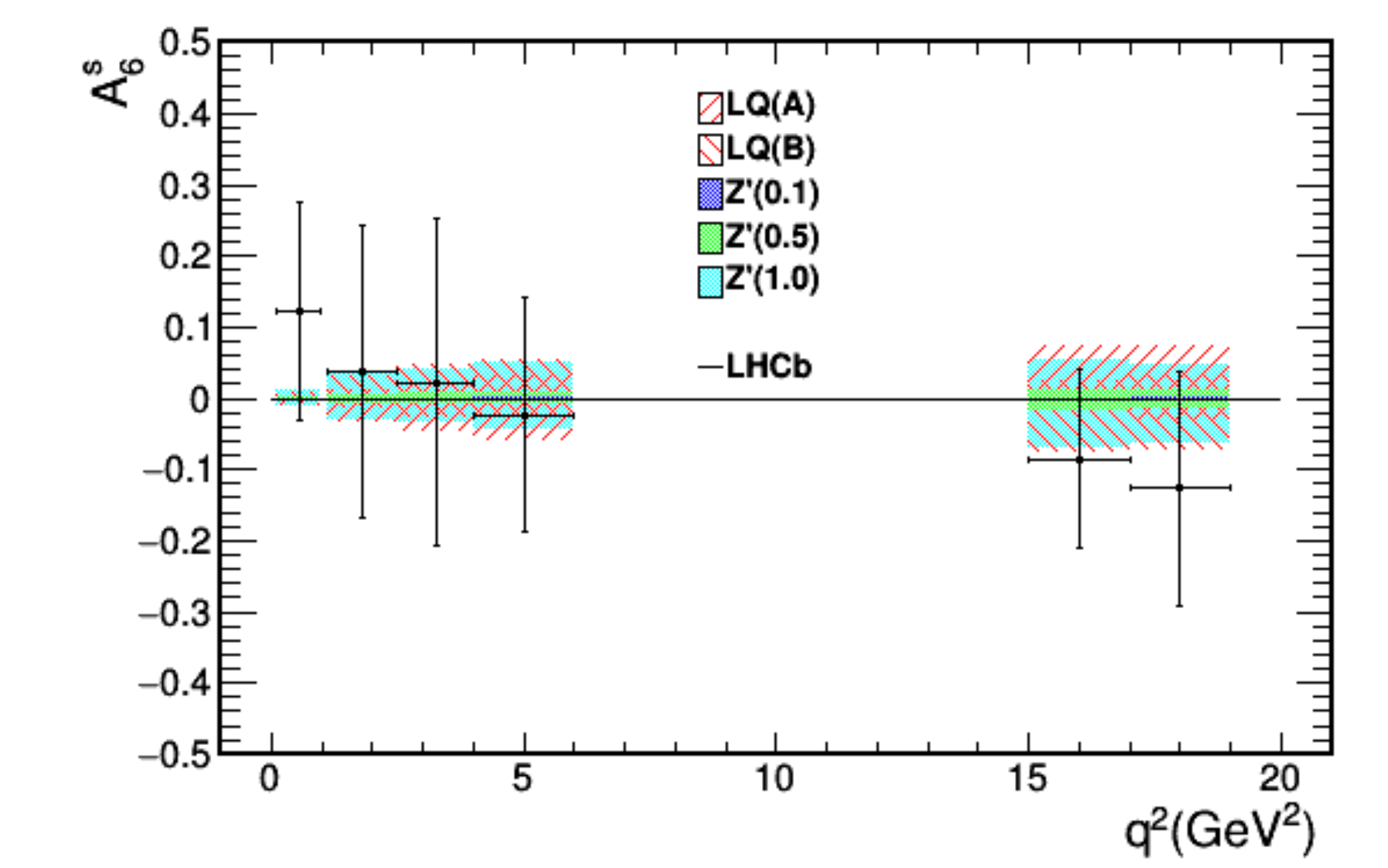}}
\vskip1truemm
{\hspace{-0.7cm}\includegraphics[height=5.3cm,width=7.95cm]{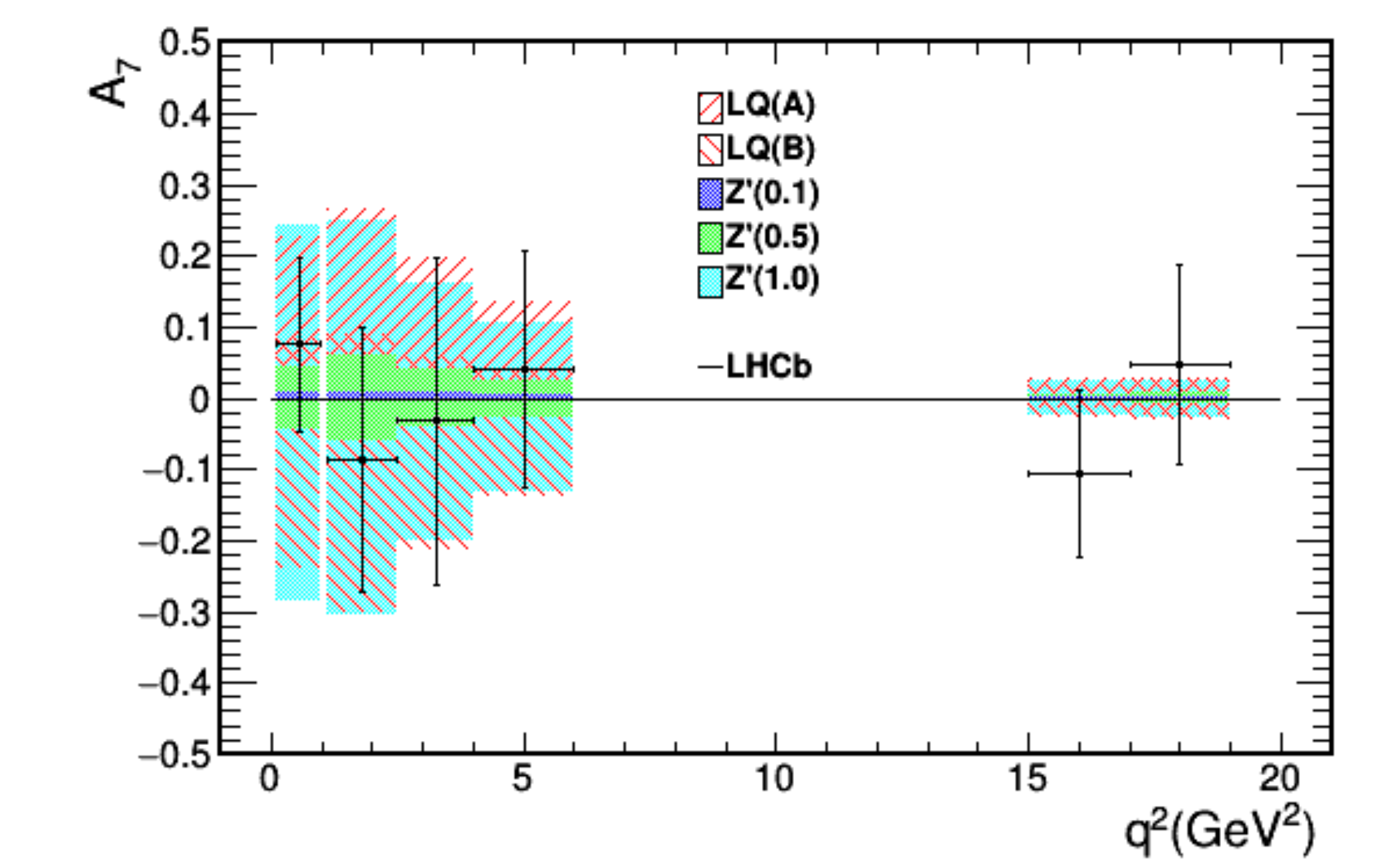}\hspace{-0.10in}
\includegraphics[height=5.3cm,width=7.95cm]{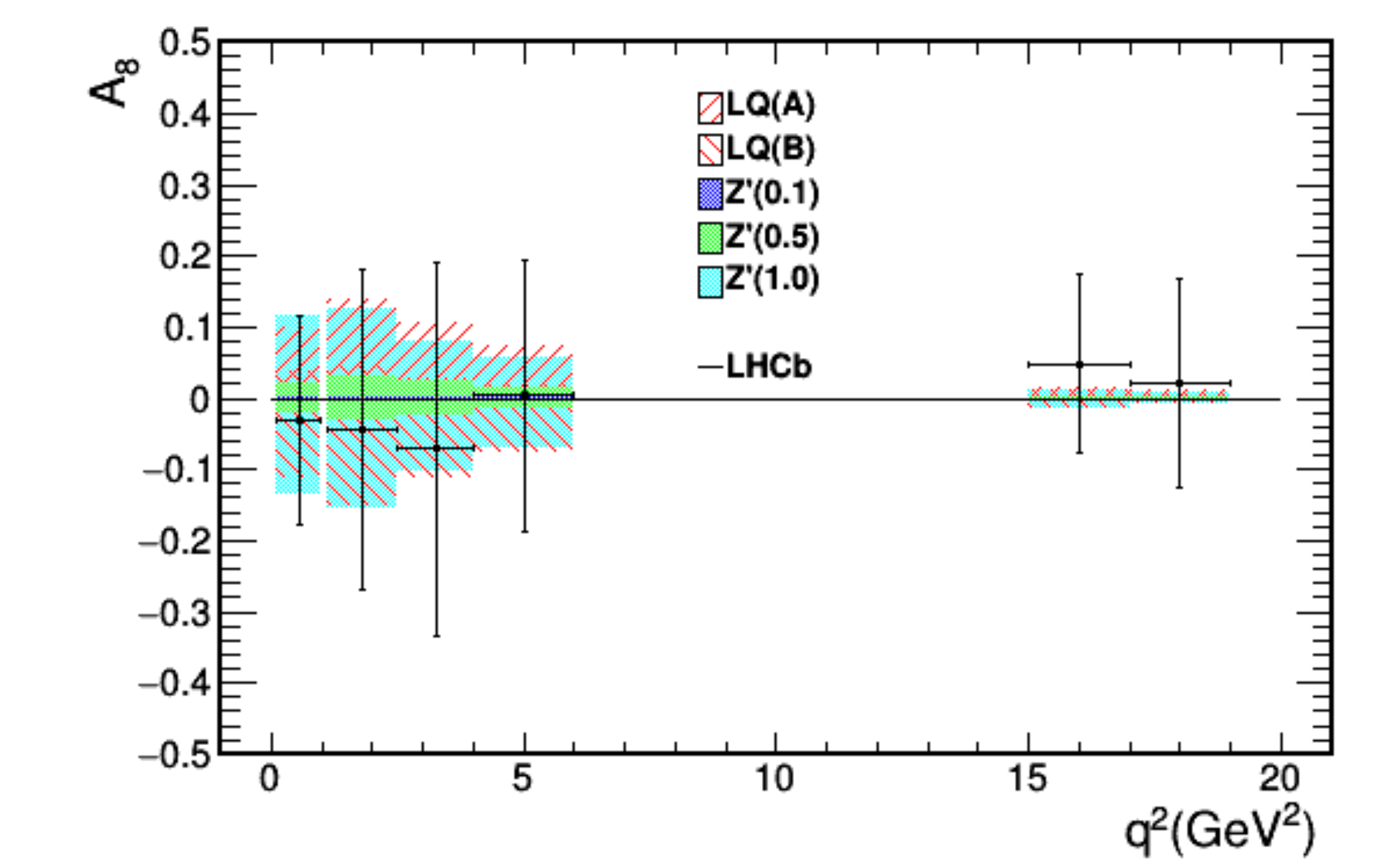}}
\vskip1truemm
{\hspace{-0.7cm}\includegraphics[height=5.3cm,width=7.95cm]{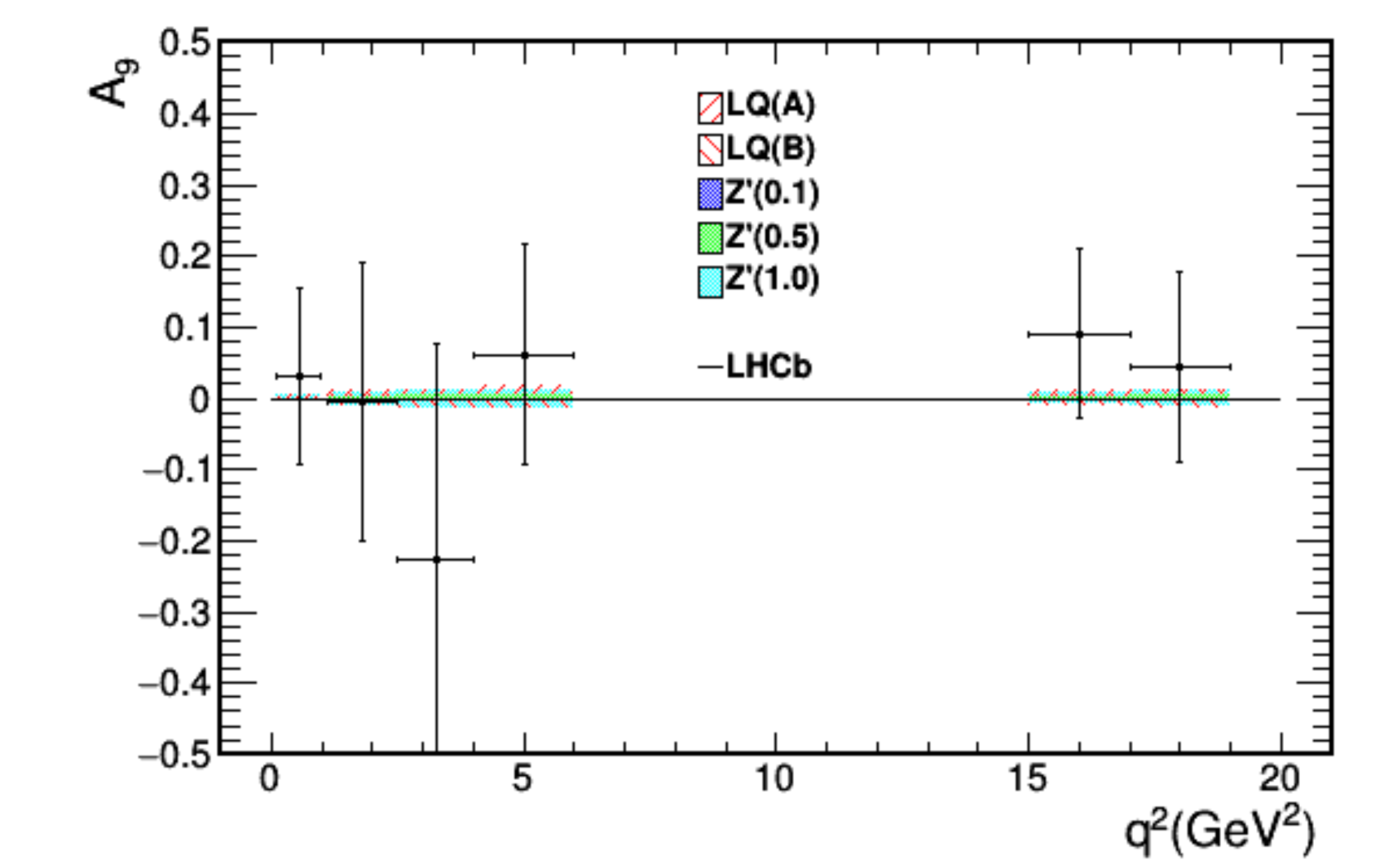}\hspace{-0.10in}
\includegraphics[height=5.3cm,width=7.95cm]{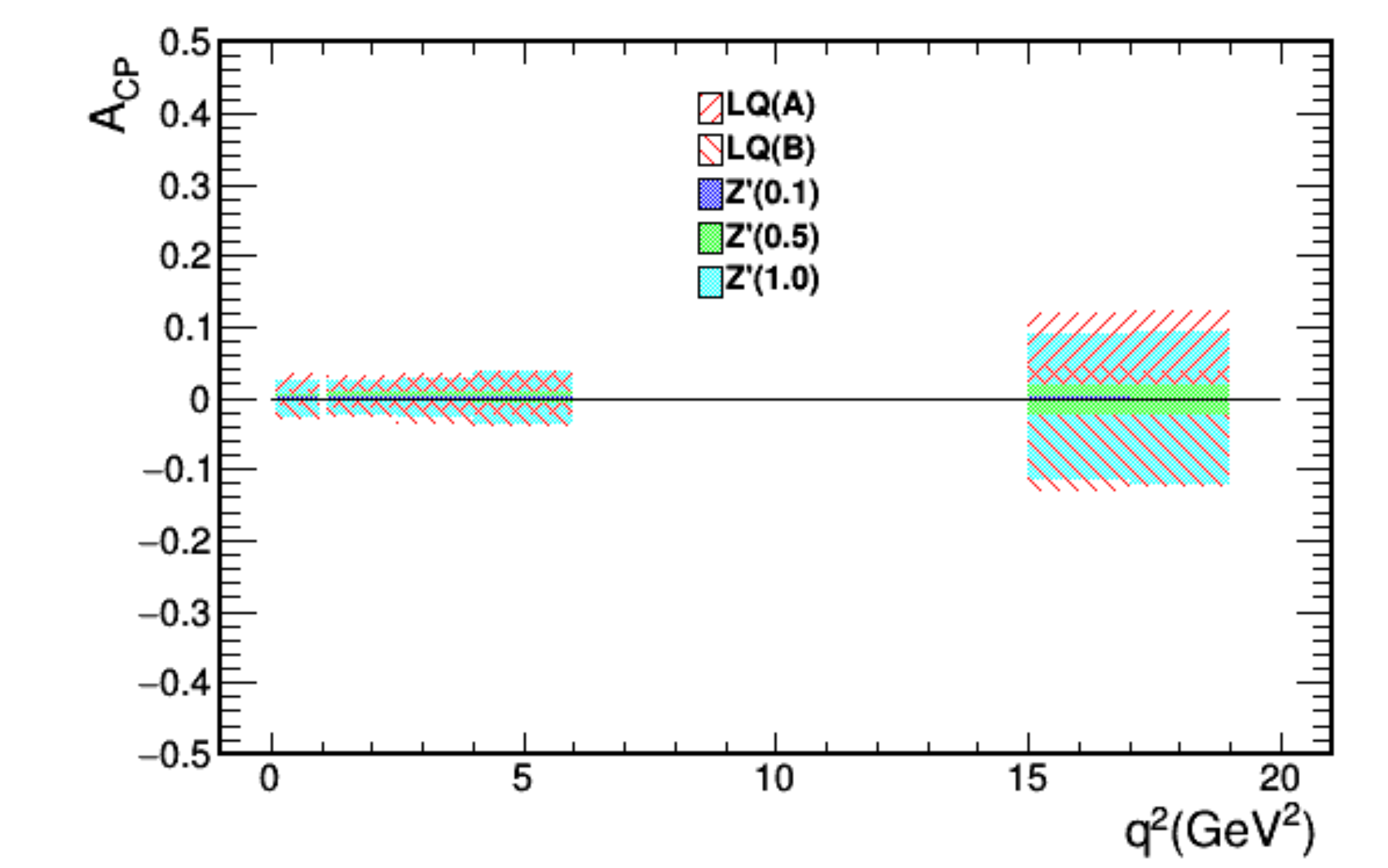}}
\caption{Predictions of the LQ model (solutions (A) and (B)) and the
  $Z'$ model with $g_L^{\mu \mu} = 0.1, 0.5, 1.0$ for the CP
  asymmetries $A_3$-$A_9$ in $B^0 \to K^{*0} \mu^+ \mu^-$ and $A_{\rm
CP}$ in $B \to K \mu^+ \mu^-$. In the models, the real
  and imaginary parts of the couplings are allowed to vary by $\pm
  2\sigma$.}
\label{LQZ'CPasym}
\end{figure}
In Fig.~\ref{LQZ'CPasym}, we present the predictions for the CP
asymmetries $A_3$-$A_9$ in $B^0 \to K^{*0} \mu^+ \mu^-$ and $A_{\rm
CP}$ in $B \to K \mu^+ \mu^-$. We consider the LQ model (solutions
(A) and (B)) and the $Z'$ model with $g_L^{\mu \mu} = 0.1, 0.5, 1.0$.
The ranges of the asymmetries are obtained by allowing the real and
imaginary parts of the couplings to vary by $\pm 2\sigma$ (taking
correlations into account). From these figures we see that
\begin{itemize}

\item The predictions of the $Z'$ model with $g_L^{\mu \mu} = 1.0$ are
  very similar to those of the LQ model in which solutions (A) and (B)
  are added.

\item Even in the presence of NP, the asymmetries $A_3$, $A_4$, $A_5$,
  and $A_9$ are very small and probably unmeasurable.

\item In the LQ and $Z'$ ($g_L^{\mu \mu} = 1.0$) models, the
  asymmetries $A_6^s$ and $A_{\rm CP}$ can approach the 10\% level in
  the high-$q^2$ region.

\item The asymmetry $A_8$ can reach 15\% in the low-$q^2$ region in
  the LQ and $Z'$ ($g_L^{\mu \mu} = 1.0$) models; it is small in the
  $Z'$ ($g_L^{\mu \mu} = 0.1, 0.5$) models.

\item The most useful asymmetry is $A_7$ in the low-$q^2$ region. In
  the LQ and $Z'$ ($g_L^{\mu \mu} = 1.0$) models, it can reach $\sim
  25\%$; in the $Z'$ ($g_L^{\mu \mu} = 0.5$) model, it can reach $\sim
  5\%$; and it is very small in the $Z'$ ($g_L^{\mu \mu} = 0.1$)
  model.

\item If a large nonzero CP asymmetry is measured, its sign
  distinguishes solutions (A) and (B) of the LQ model.

\end{itemize}
From this we see that, using CP-violating asymmetries in $B \to
K^{(*)} \mu^+ \mu^-$, it may indeed be possible to distinguish the LQ
and $Z'$ ($g_L^{\mu \mu} = 1.0$) models from $Z'$ models with
different values of $g_L^{\mu \mu}$.

Finally, it was pointed out above that the predictions of the LQ model
in which solutions (A) and (B) are added are very similar to those of
the $Z'$ model ($g_L^{\mu \mu} = 1.0$). Furthermore, we note that
these predictions are also very similar to those of the model-independent
analysis (scenario II: $C_9^{\mu\mu}({\rm NP}) = - C_{10}^{\mu\mu}({\rm
NP})$), shown in Fig.~\ref{fig:a9}. This is to be expected. Both the
model-independent and LQ fits include only $\bsmumu$ data, and for
$g_L^{\mu \mu} = 1.0$, the $Z'$ fit is dominated by the $\bsmumu$ data
(the additional constraints from $\bs$-$\bsbar$
mixing are negligible). On the other hand, in a $Z'$ model with
$g_L^{\mu \mu} < 1.0$, the constraints from $\bs$-$\bsbar$ mixing are
important, so that the predicted asymmetries are smaller than with
$g_L^{\mu \mu} = 1.0$. This is another example of how model-independent
and model-dependent fits can yield different results.

\section{Summary \& Conclusions}

There are currently a number of $B$-decay measurements involving
$\bsll$ that exhibit discrepancies with the predictions of the SM.
These include the angular analysis of $B \to K^* \mu^+\mu^-$, the
branching fraction and angular analysis of $\bs \to \phi \mu^+ \mu^-$,
and $R_K \equiv {\cal B}(B^+ \to K^+ \mu^+ \mu^-)/{\cal B}(B^+ \to K^+
e^+ e^-)$. The model-independent global analysis of
Ref.~\cite{BK*mumulatestfit1} showed that these anomalies can be
explained if there is new physics in $\bsmumu$. Assuming that the NP
Wilson coefficients are real, the four possible scenarios are (I)
$C_9^{\mu\mu}({\rm NP}) < 0$, (II) $C_9^{\mu\mu}({\rm NP}) = -
C_{10}^{\mu\mu}({\rm NP}) < 0$, (III) $C_9^{\mu\mu}({\rm NP}) = -
C_{9}^{\prime \mu\mu}({\rm NP}) < 0$, and (IV) $C_9^{\mu\mu}({\rm NP})
= - C_{10}^{\mu\mu}({\rm NP}) = -C_{9}^{\prime \mu\mu}({\rm NP}) = -
C_{10}^{\prime \mu\mu}({\rm NP}) < 0$.

Many models have been proposed as explanations of the $B$-decay
anomalies. The purpose of this paper is to investigate whether one can
distinguish among these models using measurements of CP-violating
asymmetries in $\BKstarmumu$ and $\BKmumu$. (In the SM, all
CP-violating effects are expected to be tiny.)

We begin by repeating the model-independent global analysis, this time
allowing for complex WCs. We confirm that the four scenarios I-IV do
indeed provide good fits to the data. Then, using the best-fit values
and errors of the real and imaginary parts of the WCs, we compute the
allowed ranges of the CP asymmetries in $B \to K^{(*)} \mu^+ \mu^-$.
We find that several asymmetries can be large, greater than 10\%. More
importantly, by combining the results of different CP asymmetries, it
is potentially possible to differentiate scenarios I-IV.

We then turn to a model-dependent analysis. There are two classes of
NP that can contribute to $\bsmumu$: leptoquarks and $Z'$ bosons. We
examine these two types of NP in order to determine the
characteristics of models that can explain the $B$-decay anomalies.
Note that a specific model may have additional theoretical or
experimental constraints, which must be taken into account in the
model-dependent fits. This can lead to results that are quite
different from the model-independent fits. Given a model that accounts
for the $\bsmumu$ data, we compute its predictions for CP-violating
effects. In order to generate sizeable CP asymmetries, the NP weak
phase must be large.

We consider all possible LQ models and find that three can explain the
$B$ anomalies. All have $C_9^{\mu\mu}({\rm NP}) = -
C_{10}^{\mu\mu}({\rm NP})$ (scenario II), and so are equivalent as far
as the $\bsmumu$ data are concerned. The three LQs contribute
differently to $b \to s \nu_\mu {\bar \nu}_\mu$, and so could, in
principle, be distinguished by measurements of $\bsnunubar$. However,
we find that the constraints on the models from the present
$\bsnunubar$ data are far weaker than those from $\bsmumu$, so that
the three models remain indistinguishable. That is, there is
effectively only one LQ model that can explain the $\bsmumu$
data. There are two best-fit solutions (A) and (B); both have
$|$Im(coupling)/Re(coupling)$|= O(1)$, corresponding to a large NP
weak phase.

Many $Z'$ models have been proposed to explain the $B$ anomalies, but
most of these also have $C_9^{\mu\mu}({\rm NP}) = -
C_{10}^{\mu\mu}({\rm NP})$ (scenario II). Thus, although the models
are constructed differently, all have couplings $g_L^{bs} \, {\bar s}
\gamma^{\mu} P_L b Z'_\mu$ and $g_L^{\mu \mu} \, {\bar\mu}
\gamma^{\mu} P_L \mu Z'_\mu$. $g_L^{\mu \mu}$ is necessarily real, but
$g_L^{bs}$ may be complex. The potential size of CP asymmetries is
related to the size of the weak phase of $g_L^{bs}$. The product
$g_L^{bs} g_L^{\mu \mu}$ is constrained by $\bsmumu$, while there are
constraints on $(g_L^{bs})^2$ due to the $Z'$ contribution to
$\bs$-$\bsbar$ mixing. If $g_L^{\mu \mu}$ is small, the $\bsmumu$ data
requires $g_L^{bs}$ to be large, so that the $\bs$-$\bsbar$ mixing
constraints are stringent. In particular, the measurement of
$\varphi_s^{c{\bar c}s}$, the weak phase of the mixing, constrains the
weak phase of $g_L^{bs}$ to be small. On the other hand, if $g_L^{\mu
  \mu}$ is large, $g_L^{bs}$ is small, so the $\bs$-$\bsbar$ mixing
constraints are very weak. In this case, the weak phase of $g_L^{bs}$
can be large. We therefore see that there is a whole spectrum of $Z'$
models, parametrized by the size of the $g_L^{\mu \mu}$ coupling.

We compute the predictions for the CP asymmetries in $B \to K^{(*)}
\mu^+ \mu^-$ in the LQ model (solutions (A) and (B)) and the $Z'$
model with $g_L^{\mu \mu} = 0.1, 0.5, 1.0$. We find that it may indeed
be possible to distinguish the LQ and $Z'$ models with various values
of $g_L^{\mu \mu}$ from one another. The most useful CP asymmetry is
$A_7$ in $B^0 \to K^{*0} \mu^+ \mu^-$. In the low-$q^2$ region, this
asymmetry (i) can reach $\sim 25\%$ in the LQ and $Z'$ ($g_L^{\mu \mu}
= 1.0$) models, (ii) can reach $\sim 5\%$ in the $Z'$ ($g_L^{\mu \mu}
= 0.5$) model, (iii) is very small in the $Z'$ ($g_L^{\mu \mu} = 0.1$)
model. In addition, the sign of the asymmetry distinguishes solutions
(A) and (B) of the LQ model. We therefore conclude that measurements
of CP violation in $B \to K^{(*)} \mu^+ \mu^-$ are potentially very
useful in identifying the NP responsible for the $\bsmumu$ $B$-decay
anomalies.

\bigskip
\noindent
{\bf Acknowledgements}: This work was financially supported by NSERC
of Canada (DL), by the U. S. Department of Energy under contract
DE-SC0007983 (BB). AKA and BB acknowledge the hospitality of the GPP
at the Universit\'e de Montr\'eal during the initial stages of the
work. BB thanks Alexey Petrov and Andreas Kronfeld for useful
discussions. JK would like to thank Christoph Niehoff and David Straub
for discussions and several correspondences regarding {\tt flavio}. DL
thanks Gudrun Hiller for helpful information about the CP asymmetries
$A_3$-$A_9$.


\bigskip

\appendix

{\noindent\normalfont\bfseries\Large Appendix}

This Appendix contains Tables of all $\bsmumu$ experimental data used
in the fits.

\begin{table}[htb]
\centering
\begin{tabular}{|c|c|}
\multicolumn{2}{c}{$B^0\to K^{*0}\mu^+\mu^-$ differential branching ratio}\\
 \hline
Bin (GeV$^2$) & Measurement ($\times 10^{7}$)\\
\hline
    $[0.10,0.98]$    & $1.163_{\,-0.084}^{\,+0.076} \pm {0.033}  \pm0.079$\\
    $[1.1,2.5]$        & $0.373_{\,-0.035}^{\,+0.036} \pm {0.011} \pm0.025$\\
    $[2.5,4.0]$        & $0.383_{\,-0.038}^{\,+0.035} \pm {0.010}  \pm0.026$\\
    $[4.0,6.0]$        & $0.410_{\,-0.030}^{\,+0.031}   \pm {0.011} \pm0.028$\\
    $[15.0,17.0]$    & $0.611_{\,-0.042}^{\,+0.031} \pm {0.023} \pm0.042$\\
    $[17.0,19.0]$    & $0.385_{\,-0.024}^{\,+0.029} \pm  {0.018} \pm0.026$\\
    \hline
    $[1.1,6.0]$        &  $0.392_{\,-0.019}^{\,+0.020}  \pm{0.010} \pm0.027$\\
    $[15.0,19.0]$    &  $0.488_{\,-0.022}^{\,+0.021}  \pm{0.008} \pm0.033$\\
\hline
\end{tabular}
\caption{Experimental measurements of the differential branching ratio
  of $B^0 \to K^{*0} \mu^+\mu^-$ \cite{Aaij:2016flj}. The experimental
  errors are, from left to right, statistical, systematic and due to
  the uncertainty on the $B^0 \to J/\psi K^{*0} $ and $J/\psi \to
  \mu^+ \mu^-$ branching fractions.}
\label{B0K*mumuBRmeas}
\end{table}

\begin{table}[htb]
\centering
\begin{tabular}{|c|c|c|}
\multicolumn{3}{c}{$B^0\to K^{*0}\mu^+\mu^-$ angular observables}\\
\hline
%
$q^2 \in [\,0.10\,,\,0.98\,]\,{\rm GeV}^2 $     & $q^2 \in [\,1.1\,,\,2.5\,]\,{\rm GeV}^2 $   &$q^2 \in [\,2.5\,,\,4.0\,]\,{\rm GeV}^2 $   \\
\hline
$\av{F_L}=\phantom{-}0.263\,{}^{+0.045}_{-0.044} \pm 0.017$  & $\av{F_L}=\phantom{-}0.660\,{}^{+0.083}_{-0.077} \pm 0.022$&$\av{F_L}=\phantom{-}0.876\,{}^{+0.109}_{-0.097} \pm 0.017$  \\
$\av{A_{FB}}=-0.003\,{}^{+0.058}_{-0.057} \pm 0.009$           & $\av{A_{FB}}=-0.191\,{}^{+0.068}_{-0.080} \pm 0.012$ &$\av{A_{FB}}=-0.118\,{}^{+0.082}_{-0.090} \pm 0.007$              \\
$\av{S_3}=-0.036\,{}^{+0.063}_{-0.063} \pm 0.005$            & $\av{S_3}=-0.077\,{}^{+0.087}_{-0.105} \pm 0.005$ &$\av{S_3}=\phantom{-}0.035\,{}^{+0.098}_{-0.089} \pm 0.007$          \\
$\av{S_4}=\phantom{-}0.082\,{}^{+0.068}_{-0.069} \pm 0.009$ & $\av{S_4}=-0.077\,{}^{+0.111}_{-0.113} \pm 0.005$    &$\av{S_4}=-0.234\,{}^{+0.127}_{-0.144} \pm 0.006$           \\
$\av{S_5}=\phantom{-}0.170\,{}^{+0.059}_{-0.058} \pm 0.018$  & $\av{S_5}=\phantom{-}0.137\,{}^{+0.099}_{-0.094} \pm 0.009$ & $\av{S_5}=-0.022\,{}^{+0.110}_{-0.103} \pm 0.008$   \\
$\av{S_7}=\phantom{-}0.015\,{}^{+0.059}_{-0.059} \pm 0.006$  & $\av{S_7}=-0.219\,{}^{+0.094}_{-0.104} \pm 0.004$ & $\av{S_7}=\phantom{-}0.068\,{}^{+0.120}_{-0.112} \pm 0.005$           \\
$\av{S_8}=\phantom{-}0.079\,{}^{+0.076}_{-0.075} \pm 0.007$ & $\av{S_8}=-0.098\,{}^{+0.108}_{-0.123} \pm 0.005$ & $\av{S_8}=\phantom{-}0.030\,{}^{+0.129}_{-0.131} \pm 0.006$          \\
$\av{S_9}=-0.083\,{}^{+0.058}_{-0.057} \pm 0.004$            & $\av{S_9}=-0.119\,{}^{+0.087}_{-0.104} \pm 0.005$  & $\av{S_9}=-0.092\,{}^{+0.105}_{-0.125} \pm 0.007$          \\
\hline
$q^2 \in [\,4.0\,,\,6.0\,]\,{\rm GeV}^2 $ &   $q^2 \in [\,15.0\,,\,17.0\,]\,{\rm GeV}^2 $  & $q^2 \in [\,17.0\,,\,19.0\,]\,{\rm GeV}^2 $\\
\hline
$\av{F_L}   =\phantom{-}0.611\,{}^{+0.052}_{-0.053} \pm 0.017$  &  $\av{F_L}=\phantom{-}0.349\,{}^{+0.039}_{-0.039} \pm 0.009$&$\av{F_L}   =\phantom{-}0.354\,{}^{+0.049}_{-0.048} \pm 0.025$    \\
$\av{A_{FB}}= \phantom{-}0.025\,{}^{+0.051}_{-0.052} \pm 0.004$  & $\av{A_{FB}}=\phantom{-}0.411\,{}^{+0.041}_{-0.037} \pm 0.008$ &$\av{A_{FB}}=\phantom{-}0.305\,{}^{+0.049}_{-0.048} \pm 0.013$    \\
$\av{S_3}= \phantom{-}0.035\,{}^{+0.069}_{-0.068} \pm 0.007$  & $\av{S_3}=-0.142\,{}^{+0.044}_{-0.049} \pm 0.007$    &$\av{S_3}=-0.188\,{}^{+0.074}_{-0.084} \pm 0.017$         \\
$\av{S_4}=-0.219\,{}^{+0.086}_{-0.084} \pm 0.008$  & $\av{S_4}=-0.321\,{}^{+0.055}_{-0.074} \pm 0.007$    & $\av{S_4}=-0.266\,{}^{+0.063}_{-0.072} \pm 0.010$                     \\
$\av{S_5}=-0.146\,{}^{+0.077}_{-0.078} \pm 0.011$  & $\av{S_5}=-0.316\,{}^{+0.051}_{-0.057} \pm 0.009$    & $\av{S_5}=-0.323\,{}^{+0.063}_{-0.072} \pm 0.009$               \\
$\av{S_7}=-0.016\,{}^{+0.081}_{-0.080} \pm 0.004$  & $\av{S_7}= \phantom{-}0.061\,{}^{+0.058}_{-0.058} \pm 0.005$  &$\av{S_7}=\phantom{-}0.044\,{}^{+0.073}_{-0.072} \pm 0.013$       \\
$\av{S_8}=\phantom{-}0.167\,{}^{+0.094}_{-0.091} \pm 0.004$  & $\av{S_8}=\phantom{-}0.003\,{}^{+0.061}_{-0.061} \pm 0.003$ &$\av{S_8}=\phantom{-}0.013\,{}^{+0.071}_{-0.070} \pm 0.005$   \\
$\av{S_9}=-0.032\,{}^{+0.071}_{-0.071} \pm 0.004$  & $\av{S_9}=-0.019\,{}^{+0.054}_{-0.056} \pm 0.004$ &$\av{S_9}=-0.094\,{}^{+0.065}_{-0.067} \pm 0.004$                      \\
\hline
\end{tabular}
\caption{Experimental measurements of the angular observables of
  $B^0\to K^{*0} \mu^+\mu^-$ \cite{BK*mumuLHCb2}.  The experimental
  errors are, from left to right, statistical and systematic.}
\label{tab:BtoKstar}
\end{table}

\begin{table}[htb]
\centering
\begin{tabular}{|c|c|}
\multicolumn{2}{c}{$B^+\to K^{*+}\mu^+\mu^-$ differential branching ratio}\\
\hline
Bin (GeV$^2$) & Measurement($\times 10^{9}$)\\
\hline
$[0.1-2.0]$    & $59.2 ^{+14.4}_{-13.0} \pm 4.0$   \\
$[2.0-4.0]$    & $55.9 ^{+15.9}_{-14.4} \pm 3.8 $   \\
$[4.0-6.0]$    & $24.9 ^{+11.0}_{-9.6}  \pm 1.7 $ \\
$[15.0-17.0]$   & $64.4 ^{+12.9}_{-11.5} \pm 4.4 $\\
$[17.0-22.0]$  & $11.6 ^{9.1}_{-7.6}    \pm 0.8$\\
 \hline
\end{tabular}
\caption{Experimental measurements of the differential branching ratio
  of $B^+ \to K^{*+} \mu^+\mu^-$ \cite{Aaij:2014pli}.  The experimental
  errors are, from left to right, statistical and systematic.}
\label{B+K*mumuBRmeas}
\end{table}

\begin{table}[htb]
\centering
\begin{tabular}{|c|c|}
\multicolumn{2}{c}{$B^+\to K^{+}\mu^+\mu^-$ differential branching ratio}\\
 \hline
Bin (GeV$^2$) & Measurement ($\times 10^{9}$)\\
\hline
 $[0.1-0.98]$   & $33.2 \pm 1.8 \pm 1.7$ \\
$[1.1-2.0]$    & $23.3 \pm 1.5 \pm 1.2$ \\
$[2.0-3.0]$     & $28.2 \pm 1.6 \pm 1.4$ \\
$[3.0-4.0]$    & $25.4 \pm 1.5 \pm 1.3$ \\
$[4.0-5.0]$   & $22.1 \pm 1.4 \pm 1.1$ \\
$[5.0-6.0]$     & $23.1 \pm 1.4 \pm 1.2$ \\
$[15.0-16.0]$   & $16.1 \pm 1.0 \pm 0.8$ \\
$[16.0-17.0]$    & $16.4 \pm 1.0 \pm 0.8$\\
$[17.0-18.0]$   & $20.6 \pm 1.1 \pm 1.0$\\
$[18.0-19.0]$    & $13.7 \pm 1.0 \pm 0.7$ \\
$[19.0-20.0]$   & $\phantom{0}7.4 \pm 0.8 \pm 0.4$  \\
$[20.0-21.0]$   & $\phantom{0}5.9 \pm 0.7 \pm 0.3$ \\
$[21.0-22.0]$   & $\phantom{0}4.3 \pm 0.7 \pm 0.2$  \\
\hline
$[1.1-6.0]$   & $\phantom{0}24.2 \pm 0.7 \pm 1.2$ \\
$[15.0-22.0]$   & $\phantom{0}12.1 \pm 0.4 \pm 0.6$  \\
\hline
\end{tabular}
\caption{Experimental measurements of the differential branching
  ratio of $B^+\to K^{+} \mu^+\mu^-$ \cite{Aaij:2014pli}.  The
  experimental errors are, from left to right, statistical and
  systematic.}
\label{B+KmumuBRmeas}
\end{table}

\begin{table}
\centering
\begin{tabular}{|c|c|}
\multicolumn{2}{c}{$B^0\to K^{0}\mu^+\mu^-$ differential branching ratio}\\
 \hline
Bin (GeV$^2$) & Measurement ($\times 10^{9}$)\\
\hline
$[0.1-2.0]$    & $12.2 ^{+5.9}_{-5.2} \pm 0.6$ \\
$[2.0-4.0]$    & $18.7 ^{+5.5}_{-4.9} \pm 0.9 $ \\
$[4.0-6.0]$  & $17.3 ^{+5.3}_{-4.8} \pm 0.9 $ \\
$[15.0-17.0]$  & $14.3 ^{+3.5}_{-3.2} \pm 0.7 $ \\
$[17.0-22.0]$  & $\phantom{0}7.8 ^{+1.7}_{-1.5} \pm 0.4$\\
\hline
$[1.1-6.0]$   & $18.7 ^{+3.5}_{-3.2} \pm 0.9$\\
$[15.0-22.0]$  & $ \phantom{0}9.5 ^{+1.6}_{-1.5} \pm 0.5 $ \\
\hline
\end{tabular}
\caption{Experimental measurements of the differential branching
  ratio of $B^0\to K^{0} \mu^+\mu^-$ \cite{Aaij:2014pli}.  The
  experimental errors are, from left to right, statistical and
  systematic.}
\label{B0KmumuBRmeas}
\end{table}

\begin{table}
\centering
\begin{tabular}{|c|c|}
\multicolumn{2}{c}{$\bs \to \phi \mu^+\mu^-$ differential branching ratio}\\
 \hline
Bin (GeV$^2$) & Measurement ($\times 10^{8}$)\\
\hline
$[0.1-2.0]$    & $5.85 ^{+0.73}_{-0.69} \pm{0.14} \pm{0.44}$ \\
$[2.0-5.0]$  & $2.56 ^{+0.42}_{-0.39} \pm{0.06} \pm{0.19}$ \\
$[15.0-17.0]$   & $4.52 ^{+0.57}_{-0.54} \pm{0.12} \pm{0.34}$ \\
$[17.0-19.0]$ & $3.96 ^{+0.57}_{-0.54} \pm{0.14} \pm{0.30}$ \\
\hline
\end{tabular}
\caption{Experimental measurements of the differential branching ratio
  of $\bs \to \phi \mu^+\mu^-$ \cite{BsphimumuLHCb2}. The experimental
  errors are, from left to right, statistical, systematic and due to
  the uncertainty on the branching ratio of the normalization mode
  $\bs \to J/\psi\phi$.}
\label{BsphimumuBRmeas}
\end{table}

\begin{table}[t]
\centering
\begin{tabular}{|c|c|}
\multicolumn{2}{c}{$\bs \to \phi \mu^+\mu^-$ angular observables}\\
\hline
%
$q^2 \in [\,0.1\,,\,2.0\,]\,{\rm GeV}^2 $  & $q^2 \in [\,2.0\,,\,5.0\,]\,{\rm GeV}^2 $        \\
\hline
$\av{F_L}   = \phantom{-} 0.20 ^{+0.08}_{-0.09}\pm 0.02 $      & $\av{F_L}  = \phantom{-} 0.68 ^{+ 0.16 }_{ -0.13 } \pm 0.03 $ \\
$\av{S_3}    =  -0.05 ^{+ 0.13 }_{ -0.13 } \pm 0.01 $           &  $\av{S_3}=  -0.06 ^{+ 0.19 }_{ -0.23 } \pm 0.01 $        \\
$\av{S_4}    = \phantom{-} 0.27 ^{+ 0.28 }_{ -0.18 } \pm 0.01 $ &  $\av{S_4}= -0.47 ^{+ 0.30 }_{ -0.44 } \pm 0.01 $ \\
$\av{S_7}   = \phantom{-} 0.04^{+ 0.12 }_{ -0.12 } \pm 0.00 $  &  $\av{S_7}= -0.03^{+ 0.18 }_{ -0.23 } \pm 0.01 $      \\
\hline
$q^2 \in [\,15.0\,,\,17.0\,]\,{\rm GeV}^2 $  & $q^2 \in [\,17.0\,,\,19.0\,]\,{\rm GeV}^2 $   \\
\hline
$\av{F_L}=\phantom{-} 0.23 ^{+ 0.09 }_{ -0.08 } \pm 0.02 $&$\av{F_L}= \phantom{-} 0.40 ^{+ 0.13 }_{ -0.15 } \pm 0.02 $    \\
$\av{S_3}=-0.06 ^{+ 0.16 }_{ -0.19 } \pm 0.01 $      &$\av{S_3}= -0.07 ^{+ 0.23 }_{ -0.27 } \pm 0.02 $             \\
$\av{S_4}= -0.03 ^{+ 0.15 }_{ -0.15 } \pm 0.01 $     &$\av{S_4}= -0.39 ^{+ 0.25 }_{ -0.34 } \pm 0.02 $               \\
$\av{S_7}= \phantom{-} 0. 12^{+ 0.16 }_{ -0.13 } \pm 0.01 $  &$\av{S_7}=\phantom{-} 0.20^{+ 0.29 }_{ -0.22 } \pm 0.01 $  \\
\hline
\end{tabular}
\caption{Experimental measurements of the angular observables of
  $\bs \to \phi \mu^+\mu^-$ \cite{BsphimumuLHCb2}.  The experimental
  errors are, from left to right, statistical and systematic.}
\label{Bsphimumuangmeas}
\end{table}

\begin{table}
\centering
\begin{tabular}{|c|c|}
\multicolumn{2}{c}{$B\to X_s \mu^+ \mu^-$ differential branching ratio}\\
 \hline
Bin & Measurement ($\times 10^{6}$)\\
\hline
$q^2 \in [1,6] ~{\rm GeV}^2$    & $0.66 \pm{0.88}$ \\
$q^2 > 14.2 ~{\rm GeV}^2$  & $0.60 \pm{0.31}$ \\
\hline
\end{tabular}
\caption{Experimental measurements of the differential branching ratio
  of $B\to X_s \mu^+ \mu^-$ \cite{Lees:2013nxa}.}
\label{BXsmumuBRmeas}
\end{table}

\pagebreak


\clearpage

\begin{thebibliography}{99}

\bibitem{BK*mumuLHCb1}
R.~Aaij {\it et al.} [LHCb Collaboration],
  ``Measurement of Form-Factor-Independent Observables in the Decay $B^{0} \to K^{*0} \mu^+ \mu^-$,''
  Phys.\ Rev.\ Lett.\  {\bf 111}, 191801 (2013)
  doi:10.1103/PhysRevLett.111.191801
  [arXiv:1308.1707 [hep-ex]].

\bibitem{BK*mumuLHCb2}
R.~Aaij {\it et al.} [LHCb Collaboration],
  ``Angular analysis of the $B^{0} \to K^{*0} \mu^{+} \mu^{-}$ decay using 3 fb$^{-1}$ of integrated luminosity,''
  JHEP {\bf 1602}, 104 (2016)
  doi:10.1007/JHEP02(2016)104
  [arXiv:1512.04442 [hep-ex]].

\bibitem{BK*mumuBelle}
A.~Abdesselam {\it et al.} [Belle Collaboration],
  ``Angular analysis of $B^0 \to K^\ast(892)^0 \ell^+ \ell^-$,''
  arXiv:1604.04042 [hep-ex].

\bibitem{P'5} S.~Descotes-Genon, T.~Hurth, J.~Matias and J.~Virto,
  ``Optimizing the basis of $B \to K^* l l$ observables in the full kinematic range,''
  JHEP {\bf 1305}, 137 (2013)
  doi:10.1007/JHEP05(2013)137
  [arXiv:1303.5794 [hep-ph]].

\bibitem{BK*mumuhadunc1}
S.~Descotes-Genon, L.~Hofer, J.~Matias and J.~Virto,
  ``On the impact of power corrections in the prediction of $B \to K^*\mu^+\mu^-$ observables,''
  JHEP {\bf 1412}, 125 (2014)
  doi:10.1007/JHEP12(2014)125
  [arXiv:1407.8526 [hep-ph]].

\bibitem{BK*mumuhadunc2}
J.~Lyon and R.~Zwicky,
  ``Resonances gone topsy turvy - the charm of QCD or new physics in $b \to s \ell^+ \ell^-$?,''
  arXiv:1406.0566 [hep-ph].

\bibitem{BK*mumuhadunc3}
S.~J{\"a}ger and J.~Martin Camalich,
  ``Reassessing the discovery potential of the $B \to K^{*} \ell^+\ell^-$ decays in the large-recoil region: SM challenges and BSM opportunities,''
  Phys.\ Rev.\ D {\bf 93}, 014028 (2016)
  doi:10.1103/PhysRevD.93.014028
  [arXiv:1412.3183 [hep-ph]].

\bibitem{Altmannshofer:2014rta}
  W.~Altmannshofer and D.~M.~Straub,
  ``New physics in $b\rightarrow s$ transitions after LHC run 1,''
  Eur.\ Phys.\ J.\ C {\bf 75}, no. 8, 382 (2015)
  doi:10.1140/epjc/s10052-015-3602-7
  [arXiv:1411.3161 [hep-ph]].

\bibitem{BK*mumulatestfit1}
S.~Descotes-Genon, L.~Hofer, J.~Matias and J.~Virto,
  ``Global analysis of $b\to s\ell\ell$ anomalies,''
  JHEP {\bf 1606}, 092 (2016)
  doi:10.1007/JHEP06(2016)092
  [arXiv:1510.04239 [hep-ph]].

\bibitem{BK*mumulatestfit2}
T.~Hurth, F.~Mahmoudi and S.~Neshatpour,
  ``On the anomalies in the latest LHCb data,''
  Nucl.\ Phys.\ B {\bf 909}, 737 (2016)
  doi:10.1016/j.nuclphysb.2016.05.022
  [arXiv:1603.00865 [hep-ph]].

\bibitem{BsphimumuLHCb1}
R.~Aaij {\it et al.} [LHCb Collaboration],
  ``Differential branching fraction and angular analysis of the decay $B_s^0\to\phi\mu^{+}\mu^{-}$,''
  JHEP {\bf 1307}, 084 (2013)
  doi:10.1007/JHEP07(2013)084
  [arXiv:1305.2168 [hep-ex]].

\bibitem{BsphimumuLHCb2}
R.~Aaij {\it et al.} [LHCb Collaboration],
  ``Angular analysis and differential branching fraction of the decay $B^0_s\to\phi\mu^+\mu^-$,''
  JHEP {\bf 1509}, 179 (2015)
  doi:10.1007/JHEP09(2015)179
  [arXiv:1506.08777 [hep-ex]].

\bibitem{latticeQCD1}
R.~R.~Horgan, Z.~Liu, S.~Meinel and M.~Wingate,
  ``Calculation of $B^0 \to K^{*0} \mu^+ \mu^-$ and $B_s^0 \to \phi \mu^+ \mu^-$ observables using form factors from lattice QCD,''
  Phys.\ Rev.\ Lett.\  {\bf 112}, 212003 (2014)
  doi:10.1103/PhysRevLett.112.212003
  [arXiv:1310.3887 [hep-ph]],

\bibitem{latticeQCD2}
  ``Rare $B$ decays using lattice QCD form factors,''
  PoS LATTICE {\bf 2014}, 372 (2015)
  [arXiv:1501.00367 [hep-lat]].

\bibitem{QCDsumrules}
A.~Bharucha, D.~M.~Straub and R.~Zwicky,
  ``$B\to V\ell^+\ell^-$ in the Standard Model from light-cone sum rules,''
  JHEP {\bf 1608}, 098 (2016)
  doi:10.1007/JHEP08(2016)098
  [arXiv:1503.05534 [hep-ph]].

\bibitem{RKexpt} R.~Aaij {\it et al.}  [LHCb Collaboration],
  ``Test of lepton universality using $B^{+}\rightarrow K^{+}\ell^{+}\ell^{-}$ decays,''
  Phys.\ Rev.\ Lett.\  {\bf 113}, 151601 (2014)
  [arXiv:1406.6482 [hep-ex]].

\bibitem{IsidoriRK}
M.~Bordone, G.~Isidori and A.~Pattori,
  ``On the Standard Model predictions for $R_K$ and $R_{K^*}$,''
  Eur.\ Phys.\ J.\ C {\bf 76}, no. 8, 440 (2016)
  doi:10.1140/epjc/s10052-016-4274-7
  [arXiv:1605.07633 [hep-ph]].

\bibitem{bsmumuCPC}
A.~K.~Alok, A.~Datta, A.~Dighe, M.~Duraisamy, D.~Ghosh and D.~London,
  ``New Physics in $b \to s \mu^+ \mu^-$: CP-Conserving Observables,''
  JHEP {\bf 1111}, 121 (2011)
  doi:10.1007/JHEP11(2011)121
  [arXiv:1008.2367 [hep-ph]].

\bibitem{bsmumuCPV}
A.~K.~Alok, A.~Datta, A.~Dighe, M.~Duraisamy, D.~Ghosh and D.~London,
  ``New Physics in $b \to s \mu^+ \mu^-$: CP-Violating Observables,''
  JHEP {\bf 1111}, 122 (2011)
  doi:10.1007/JHEP11(2011)122
  [arXiv:1103.5344 [hep-ph]].

\bibitem{Descotes-Genon:2013wba}
  S.~Descotes-Genon, J.~Matias and J.~Virto,
  ``Understanding the $B\to K^*\mu^+\mu^-$ Anomaly,''
  Phys.\ Rev.\ D {\bf 88}, 074002 (2013)
  doi:10.1103/PhysRevD.88.074002
  [arXiv:1307.5683 [hep-ph]].

\bibitem{AlexLenz}
S.~J\"ager, K.~Leslie, M.~Kirk and A.~Lenz,
  ``Charming new physics in rare B-decays and mixing?,''
  arXiv:1701.09183 [hep-ph].

\bibitem{CCO}
L.~Calibbi, A.~Crivellin and T.~Ota,
  ``Effective Field Theory Approach to $b\to s \ell\ell^{(\prime)}$, $B\to K^{(*)} \nu {\bar\nu}$ and $B \to D^{(*)} \tau\nu$ with Third Generation Couplings,''
  Phys.\ Rev.\ Lett.\  {\bf 115}, 181801 (2015)
  doi:10.1103/PhysRevLett.115.181801
  [arXiv:1506.02661 [hep-ph]].

\bibitem{AGC}
R.~Alonso, B.~Grinstein and J.~Martin Camalich,
  ``Lepton universality violation and lepton flavor conservation in $B$-meson decays,''
  JHEP {\bf 1510}, 184 (2015)
  doi:10.1007/JHEP10(2015)184
  [arXiv:1505.05164 [hep-ph]].

\bibitem{HS1}
  G.~Hiller and M.~Schmaltz,
  ``$R_K$ and future $b \to s \ell \ell$ BSM opportunities,''
  Phys.\ Rev.\ D {\bf 90} (2014) 054014
  [arXiv:1408.1627 [hep-ph]].

\bibitem{GNR}
B.~Gripaios, M.~Nardecchia and S.~A.~Renner,
  ``Composite leptoquarks and anomalies in $B$-meson decays,''
  JHEP {\bf 1505}, 006 (2015)
  doi:10.1007/JHEP05(2015)006
  [arXiv:1412.1791 [hep-ph]].

\bibitem{VH}
I.~de Medeiros Varzielas and G.~Hiller,
  ``Clues for flavor from rare lepton and quark decays,''
  JHEP {\bf 1506}, 072 (2015)
  doi:10.1007/JHEP06(2015)072
  [arXiv:1503.01084 [hep-ph]].

\bibitem{SM}
S.~Sahoo and R.~Mohanta,
  ``Scalar leptoquarks and the rare $B$ meson decays,''
  Phys.\ Rev.\ D {\bf 91}, no. 9, 094019 (2015)
  doi:10.1103/PhysRevD.91.094019
  [arXiv:1501.05193 [hep-ph]].

\bibitem{FK}
S.~Fajfer and N.~Ko{\v s}nik,
  ``Vector leptoquark resolution of $R_K$ and $R_{D^{(*)}}$ puzzles,''
  Phys.\ Lett.\ B {\bf 755}, 270 (2016)
  doi:10.1016/j.physletb.2016.02.018
  [arXiv:1511.06024 [hep-ph]].

\bibitem{BFK}
D.~Be{\v c}irevi{\' c}, S.~Fajfer and N.~Ko{\v s}nik,
  ``Lepton flavor nonuniversality in $b \to s \ell^+\ell^-$ processes,''
  Phys.\ Rev.\ D {\bf 92}, no. 1, 014016 (2015)
  doi:10.1103/PhysRevD.92.014016
  [arXiv:1503.09024 [hep-ph]].

\bibitem{BKSZ}
D.~Be{\v c}irevi{\' c}, N.~Ko{\v s}nik, O.~Sumensari and R.~Zukanovich Funchal,
  ``Palatable Leptoquark Scenarios for Lepton Flavor Violation in Exclusive $b\to s\ell_1\ell_2$ modes,''
  JHEP {\bf 1611}, 035 (2016)
  doi:10.1007/JHEP11(2016)035
  [arXiv:1608.07583 [hep-ph]].

\bibitem{Crivellin:2015lwa}
  A.~Crivellin, G.~D'Ambrosio and J.~Heeck,
  ``Addressing the LHC flavor anomalies with horizontal gauge symmetries,''
  Phys.\ Rev.\ D {\bf 91}, 075006 (2015)
  doi:10.1103/PhysRevD.91.075006
  [arXiv:1503.03477 [hep-ph]].

\bibitem{Isidori} A.~Greljo, G.~Isidori and D.~Marzocca,
  ``On the breaking of Lepton Flavor Universality in B decays,''
  JHEP {\bf 1507}, 142 (2015)
  doi:10.1007/JHEP07(2015)142
  [arXiv:1506.01705 [hep-ph]].

\bibitem{dark}
  D.~Aristizabal Sierra, F.~Staub and A.~Vicente,
  ``Shedding light on the $b\to s$ anomalies with a dark sector,''
  Phys.\ Rev.\ D {\bf 92}, 015001 (2015)
  doi:10.1103/PhysRevD.92.015001
  [arXiv:1503.06077 [hep-ph]].

\bibitem{Chiang}
C.~W.~Chiang, X.~G.~He and G.~Valencia,
``$Z'$ model for $b \to s \ell {\bar\ell}$ flavor anomalies,''
  Phys.\ Rev.\ D {\bf 93}, 074003 (2016)
  doi:10.1103/PhysRevD.93.074003
  [arXiv:1601.07328 [hep-ph]].

\bibitem{Virto}
S.~M.~Boucenna, A.~Celis, J.~Fuentes-Martin, A.~Vicente and J.~Virto,
  ``Non-abelian gauge extensions for $B$-decay anomalies,''
  Phys.\ Lett.\ B {\bf 760}, 214 (2016)
  doi:10.1016/j.physletb.2016.06.067
  [arXiv:1604.03088 [hep-ph]],
  ``Phenomenology of an $SU(2) \times SU(2) \times U(1)$ model with lepton-flavour non-universality,''
  JHEP {\bf 1612}, 059 (2016)
  doi:10.1007/JHEP12(2016)059
  [arXiv:1608.01349 [hep-ph]].

\bibitem{GGH}
R.~Gauld, F.~Goertz and U.~Haisch,
  ``On minimal $Z'$ explanations of the $B\to K^*\mu^+\mu^-$ anomaly,''
  Phys.\ Rev.\ D {\bf 89}, 015005 (2014)
  doi:10.1103/PhysRevD.89.015005
  [arXiv:1308.1959 [hep-ph]],
  ``An explicit $Z'$-boson explanation of the $B \to K^* \mu^+ \mu^-$ anomaly,''
  JHEP {\bf 1401}, 069 (2014)
  doi:10.1007/JHEP01(2014)069
  [arXiv:1310.1082 [hep-ph]].

\bibitem{BG}
A.~J.~Buras and J.~Girrbach,
  ``Left-handed $Z'$ and $Z$ FCNC quark couplings facing new $\bsmumu$ data,''
  JHEP {\bf 1312}, 009 (2013)
  doi:10.1007/JHEP12(2013)009
  [arXiv:1309.2466 [hep-ph]].

\bibitem{BFG}
A.~J.~Buras, F.~De Fazio and J.~Girrbach,
  ``331 models facing new $b \to s\mu^+ \mu^-$ data,''
  JHEP {\bf 1402}, 112 (2014)
  doi:10.1007/JHEP02(2014)112
  [arXiv:1311.6729 [hep-ph]].

\bibitem{Perimeter}
W.~Altmannshofer, S.~Gori, M.~Pospelov and I.~Yavin,
  ``Quark flavor transitions in $L_\mu-L_\tau$ models,''
  Phys.\ Rev.\ D {\bf 89}, 095033 (2014)
  doi:10.1103/PhysRevD.89.095033
  [arXiv:1403.1269 [hep-ph]].

\bibitem{CDH}
A.~Crivellin, G.~D'Ambrosio and J.~Heeck,
  ``Explaining $h\to\mu^\pm\tau^\mp$, $B\to K^* \mu^+\mu^-$ and $B\to K \mu^+\mu^-/B\to K e^+e^-$ in a two-Higgs-doublet model with gauged $L_\mu-L_\tau$,''
  Phys.\ Rev.\ Lett.\  {\bf 114}, 151801 (2015)
  doi:10.1103/PhysRevLett.114.151801
  [arXiv:1501.00993 [hep-ph]],
``Addressing the LHC flavor anomalies with horizontal gauge symmetries,''
  Phys.\ Rev.\ D {\bf 91}, no. 7, 075006 (2015)
  doi:10.1103/PhysRevD.91.075006
  [arXiv:1503.03477 [hep-ph]].

\bibitem{SSV}
D.~Aristizabal Sierra, F.~Staub and A.~Vicente,
  ``Shedding light on the $b\to s$ anomalies with a dark sector,''
  Phys.\ Rev.\ D {\bf 92}, no. 1, 015001 (2015)
  doi:10.1103/PhysRevD.92.015001
  [arXiv:1503.06077 [hep-ph]].

\bibitem{CHMNPR}
A.~Crivellin, L.~Hofer, J.~Matias, U.~Nierste, S.~Pokorski and J.~Rosiek,
  ``Lepton-flavour violating $B$ decays in generic $Z'$ models,''
  Phys.\ Rev.\ D {\bf 92}, no. 5, 054013 (2015)
  doi:10.1103/PhysRevD.92.054013
  [arXiv:1504.07928 [hep-ph]].

\bibitem{CMJS}
A.~Celis, J.~Fuentes-Martin, M.~Jung and H.~Serodio,
  ``Family nonuniversal $Z'$ models with protected flavor-changing interactions,''
  Phys.\ Rev.\ D {\bf 92}, no. 1, 015007 (2015)
  doi:10.1103/PhysRevD.92.015007
  [arXiv:1505.03079 [hep-ph]].

\bibitem{BDW}
G.~B\'elanger, C.~Delaunay and S.~Westhoff,
  ``A Dark Matter Relic From Muon Anomalies,''
  Phys.\ Rev.\ D {\bf 92}, 055021 (2015)
  doi:10.1103/PhysRevD.92.055021
  [arXiv:1507.06660 [hep-ph]].

\bibitem{FNZ}
A.~Falkowski, M.~Nardecchia and R.~Ziegler,
  ``Lepton Flavor Non-Universality in $B$-meson Decays from a $U(2)$ Flavor Model,''
  JHEP {\bf 1511}, 173 (2015)
  doi:10.1007/JHEP11(2015)173
  [arXiv:1509.01249 [hep-ph]].

\bibitem{AQSS}
B.~Allanach, F.~S.~Queiroz, A.~Strumia and S.~Sun,
  ``$Z'$ models for the LHCb and $g-2$ muon anomalies,''
  Phys.\ Rev.\ D {\bf 93}, no. 5, 055045 (2016)
  doi:10.1103/PhysRevD.93.055045
  [arXiv:1511.07447 [hep-ph]].

\bibitem{CFL}
A.~Celis, W.~Z.~Feng and D.~L\"ust,
  ``Stringy explanation of $b\to s\ell^+ \ell^-$ anomalies,''
  JHEP {\bf 1602}, 007 (2016)
  doi:10.1007/JHEP02(2016)007
  [arXiv:1512.02218 [hep-ph]].

\bibitem{Hou}
K.~Fuyuto, W.~S.~Hou and M.~Kohda,
  ``$Z'$-induced FCNC decays of top, beauty, and strange quarks,''
  Phys.\ Rev.\ D {\bf 93}, no. 5, 054021 (2016)
  doi:10.1103/PhysRevD.93.054021
  [arXiv:1512.09026 [hep-ph]].

\bibitem{CHV}
C.~W.~Chiang, X.~G.~He and G.~Valencia,
  ``$Z'$ model for $b\to s \ell {\bar\ell}$ flavor anomalies,''
  Phys.\ Rev.\ D {\bf 93}, no. 7, 074003 (2016)
  doi:10.1103/PhysRevD.93.074003
  [arXiv:1601.07328 [hep-ph]].

\bibitem{CFV}
  A.~Celis, W.~Z.~Feng and M.~Vollmann,
  Phys.\ Rev.\ D {\bf 95}, no. 3, 035018 (2017)
  doi:10.1103/PhysRevD.95.035018
  [arXiv:1608.03894 [hep-ph]].

\bibitem{CFGI}
A.~Crivellin, J.~Fuentes-Martin, A.~Greljo and G.~Isidori,
  Phys.\ Lett.\ B {\bf 766}, 77 (2017)
  doi:10.1016/j.physletb.2016.12.057
  [arXiv:1611.02703 [hep-ph]].

\bibitem{IGG}
I.~Garcia Garcia,
  JHEP {\bf 1703}, 040 (2017)
  doi:10.1007/JHEP03(2017)040
  [arXiv:1611.03507 [hep-ph]].

\bibitem{BdecaysDM}
J.~M.~Cline, J.~M.~Cornell, D.~London and R.~Watanabe,
  Phys.\ Rev.\ D {\bf 95}, no. 9, 095015 (2017)
  doi:10.1103/PhysRevD.95.095015
  [arXiv:1702.00395 [hep-ph]].

\bibitem{Bhatia:2017tgo}
 D.~Bhatia, S.~Chakraborty and A.~Dighe,
  JHEP {\bf 1703}, 117 (2017)
  doi:10.1007/JHEP03(2017)117
  [arXiv:1701.05825 [hep-ph]].

\bibitem{RKRDmodels}
 B.~Bhattacharya, A.~Datta, J.~P.~Guévin, D.~London and R.~Watanabe,
  JHEP {\bf 1701}, 015 (2017)
  doi:10.1007/JHEP01(2017)015
  [arXiv:1609.09078 [hep-ph]].

\bibitem{BHP}
  C.~Bobeth, G.~Hiller and G.~Piranishvili,
  ``CP Asymmetries in ${\bar B} \to \bar{K}^* (\to \bar{K} \pi) \bar{\ell} \ell$ and Untagged $\bar{B}_s$, $B_s \to \phi (\to K^{+} K^-) \bar{\ell} \ell$ Decays at NLO,''
  JHEP {\bf 0807}, 106 (2008)
  doi:10.1088/1126-6708/2008/07/106
  [arXiv:0805.2525 [hep-ph]].

\bibitem{BK*mumuCPV}
W.~Altmannshofer, P.~Ball, A.~Bharucha, A.~J.~Buras, D.~M.~Straub and M.~Wick,
  ``Symmetries and Asymmetries of $B \to K^{*} \mu^{+} \mu^{-}$ Decays in the Standard Model and Beyond,''
  JHEP {\bf 0901}, 019 (2009)
  doi:10.1088/1126-6708/2009/01/019
  [arXiv:0811.1214 [hep-ph]].

\bibitem{flavio}
David Straub, \textit{flavio v0.11, 2016.}
  \href{http://dx.doi.org/10.5281/zenodo.59840}{http://dx.doi.org/10.5281/zenodo.59840}

\bibitem{James:1975dr}
  F.~James and M.~Roos,
  ``Minuit: A System for Function Minimization and Analysis of the Parameter Errors and Correlations,''
  Comput.\ Phys.\ Commun.\  {\bf 10}, 343 (1975).
  doi:10.1016/0010-4655(75)90039-9

\bibitem{James:2004xla}
  F.~James and M.~Winkler,
  ``MINUIT User's Guide,''
  http://inspirehep.net/record/1258345?ln=en

\bibitem{James:1994vla}
  F.~James,
  ``MINUIT Function Minimization and Error Analysis:  Reference Manual Version 94.1,''
  CERN-D-506, CERN-D506.

\bibitem{pdg}
C.~Patrignani {\it et al.} [Particle Data Group],
  ``Review of Particle Physics,''
  Chin.\ Phys.\ C {\bf 40}, no. 10, 100001 (2016).
  doi:10.1088/1674-1137/40/10/100001

\bibitem{Aaij:2014pli}
R.~Aaij {\it et al.} [LHCb Collaboration],
  ``Differential branching fractions and isospin asymmetries of $B \to K^{(*)} \mu^+ \mu^-$ decays,''
  JHEP {\bf 1406}, 133 (2014)
  doi:10.1007/JHEP06(2014)133
  [arXiv:1403.8044 [hep-ex]].

\bibitem{Lees:2013nxa}
J.~P.~Lees {\it et al.} [BaBar Collaboration],
  ``Measurement of the $B \to X_s l^+l^-$ branching fraction and search for direct CP violation from a sum of exclusive final states,''
  Phys.\ Rev.\ Lett.\  {\bf 112}, 211802 (2014)
  doi:10.1103/PhysRevLett.112.211802
  [arXiv:1312.5364 [hep-ex]].

\bibitem{Aaij:2013aka}
R.~Aaij {\it et al.} [LHCb Collaboration],
  ``Measurement of the $B^0_s \to \mu^+ \mu^-$ branching fraction and search for $B^0 \to \mu^+ \mu^-$ decays at the LHCb experiment,''
  Phys.\ Rev.\ Lett.\  {\bf 111}, 101805 (2013)
  doi:10.1103/PhysRevLett.111.101805
  [arXiv:1307.5024 [hep-ex]].

\bibitem{CMS:2014xfa}
V.~Khachatryan {\it et al.} [CMS and LHCb Collaborations],
  ``Observation of the rare $B^0_s\to\mu^+\mu^-$ decay from the combined analysis of CMS and LHCb data,''
  Nature {\bf 522}, 68 (2015)
  doi:10.1038/nature14474
  [arXiv:1411.4413 [hep-ex]].

\bibitem{Bailey:2015dka}
  J.~A.~Bailey {\it et al.},
  ``$B\to Kl^+l^-$ decay form factors from three-flavor lattice QCD,''
  Phys.\ Rev.\ D {\bf 93}, no. 2, 025026 (2016)
  doi:10.1103/PhysRevD.93.025026
  [arXiv:1509.06235 [hep-lat]].

\bibitem{Descotes-Genon:2015hea}
  S.~Descotes-Genon and J.~Virto,
  ``Time dependence in $B \to V\ell\ell$ decays,''
  JHEP {\bf 1504}, 045 (2015)
  Erratum: [JHEP {\bf 1507}, 049 (2015)]
  doi:10.1007/JHEP04(2015)045, 10.1007/JHEP07(2015)049
  [arXiv:1502.05509 [hep-ph]].

\bibitem{Asatryan:2002iy}
  H.~H.~Asatryan, H.~M.~Asatrian, C.~Greub and M.~Walker,
  ``Complete gluon bremsstrahlung corrections to the process $b \to  s l^+ l^-$,''
  Phys.\ Rev.\ D {\bf 66}, 034009 (2002)
  doi:10.1103/PhysRevD.66.034009
  [hep-ph/0204341].

\bibitem{Ghinculov:2003qd}
  A.~Ghinculov, T.~Hurth, G.~Isidori and Y.~P.~Yao,
  ``The Rare decay $B \to  X_s l^+ l^-$ to NNLL precision for arbitrary dilepton invariant mass,''
  Nucl.\ Phys.\ B {\bf 685}, 351 (2004)
  doi:10.1016/j.nuclphysb.2004.02.028
  [hep-ph/0312128].

\bibitem{Huber:2005ig}
  T.~Huber, E.~Lunghi, M.~Misiak and D.~Wyler,
  ``Electromagnetic logarithms in $\bar B \to  X_s l^+ l^-$,''
  Nucl.\ Phys.\ B {\bf 740}, 105 (2006)
  doi:10.1016/j.nuclphysb.2006.01.037
  [hep-ph/0512066].

\bibitem{Huber:2007vv}
  T.~Huber, T.~Hurth and E.~Lunghi,
  ``Logarithmically Enhanced Corrections to the Decay Rate and Forward Backward Asymmetry in $\bar{B} \to X_s \ell^+ \ell^-$,''
  Nucl.\ Phys.\ B {\bf 802}, 40 (2008)
  doi:10.1016/j.nuclphysb.2008.04.028
  [arXiv:0712.3009 [hep-ph]].

\bibitem{DatLon}
A.~Datta and D.~London,
  ``Measuring new physics parameters in $B$ penguin decays,''
  Phys.\ Lett.\ B {\bf 595}, 453 (2004)
  doi:10.1016/j.physletb.2004.06.069
  [hep-ph/0404130].

\bibitem{Sakakietal}
Y.~Sakaki, M.~Tanaka, A.~Tayduganov and R.~Watanabe,
  ``Testing leptoquark models in $\bar B \to D^{(*)} \tau \bar\nu$,''
  Phys.\ Rev.\ D {\bf 88}, no. 9, 094012 (2013)
  doi:10.1103/PhysRevD.88.094012
  [arXiv:1309.0301 [hep-ph]].

\bibitem{Buras:2014fpa}
  A.~J.~Buras, J.~Girrbach-Noe, C.~Niehoff and D.~M.~Straub,
  ``$ B\to {K}^{\left(\ast \right)}\nu \overline{\nu} $ decays in the Standard Model and beyond,''
  JHEP {\bf 1502}, 184 (2015)
  doi:10.1007/JHEP02(2015)184
  [arXiv:1409.4557 [hep-ph]].

\bibitem{Grygier:2017tzo}
  J.~Grygier {\it et al.} [Belle Collaboration],
  ``Search for $\boldsymbol{B\to h\nu\bar{\nu}}$ decays with semileptonic tagging at Belle,''
  arXiv:1702.03224 [hep-ex].

\bibitem{RKRD}
B.~Bhattacharya, A.~Datta, D.~London and S.~Shivashankara,
  ``Simultaneous Explanation of the $R_K$ and $R(D^{(*)})$ Puzzles,''
  Phys.\ Lett.\ B {\bf 742}, 370 (2015)
  [arXiv:1412.7164 [hep-ph]].

\bibitem{RD_BaBar}
  J.~P.~Lees {\it et al.} [BaBar Collaboration],
  ``Measurement of an Excess of $\bar{B} \to D^{(*)}\tau^- \bar{\nu}_\tau$ Decays and Implications for Charged Higgs Bosons,''
  Phys.\ Rev.\ D {\bf 88}, 072012 (2013)
  doi:10.1103/PhysRevD.88.072012
  [arXiv:1303.0571 [hep-ex]].

\bibitem{RD_Belle}
M.~Huschle {\it et al.} [Belle Collaboration],
  ``Measurement of the branching ratio of $\bar{B} \to D^{(\ast)} \tau^- \bar{\nu}_\tau$ relative to $\bar{B} \to D^{(\ast)} \ell^- \bar{\nu}_\ell$ decays with hadronic tagging at Belle,''
  Phys.\ Rev.\ D {\bf 92}, 072014 (2015)
  doi:10.1103/PhysRevD.92.072014
  [arXiv:1507.03233 [hep-ex]].

\bibitem{RD_LHCb}
R.~Aaij {\it et al.} [LHCb Collaboration],
  ``Measurement of the ratio of branching fractions $\mathcal{B}(\bar{B}^0 \to D^{*+}\tau^{-}\bar{\nu}_{\tau})/\mathcal{B}(\bar{B}^0 \to D^{*+}\mu^{-}\bar{\nu}_{\mu})$,''
  Phys.\ Rev.\ Lett.\  {\bf 115}, 111803 (2015)
  Addendum: [Phys.\ Rev.\ Lett.\  {\bf 115}, 159901 (2015)]
  doi:10.1103/PhysRevLett.115.159901, 10.1103/PhysRevLett.115.111803
  [arXiv:1506.08614 [hep-ex]].

\bibitem{Buchalla:1995vs}
G.~Buchalla, A.~J.~Buras and M.~E.~Lautenbacher,
  ``Weak decays beyond leading logarithms,''
  Rev.\ Mod.\ Phys.\  {\bf 68}, 1125 (1996)
  doi:10.1103/RevModPhys.68.1125
  [hep-ph/9512380].

\bibitem{HFAG}
Y.~Amhis {\it et al.} [Heavy Flavor Averaging Group (HFAG) Collaboration],
  ``Averages of $b$-hadron, $c$-hadron, and $\tau$-lepton properties as of summer 2014,''
  arXiv:1412.7515 [hep-ex].

\bibitem{Gamiz:2009ku}
  E.~Gamiz {\it et al.} [HPQCD Collaboration],
  ``Neutral $B$ Meson Mixing in Unquenched Lattice QCD,''
  Phys.\ Rev.\ D {\bf 80}, 014503 (2009)
  doi:10.1103/PhysRevD.80.014503
  [arXiv:0902.1815 [hep-lat]].

\bibitem{Aoki:2014nga}
  Y.~Aoki, T.~Ishikawa, T.~Izubuchi, C.~Lehner and A.~Soni,
  ``Neutral $B$ meson mixings and $B$ meson decay constants with static heavy and domain-wall light quarks,''
  Phys.\ Rev.\ D {\bf 91}, no. 11, 114505 (2015)
  doi:10.1103/PhysRevD.91.114505
  [arXiv:1406.6192 [hep-lat]].

\bibitem{Aoki:2016frl}
  S.~Aoki {\it et al.},
  ``Review of lattice results concerning low-energy particle physics,''
  Eur.\ Phys.\ J.\ C {\bf 77}, no. 2, 112 (2017)
  doi:10.1140/epjc/s10052-016-4509-7
  [arXiv:1607.00299 [hep-lat]].

\bibitem{Charles:2004jd}
  J.~Charles {\it et al.} [CKMfitter Group],
  ``CP violation and the CKM matrix: Assessing the impact of the asymmetric $B$ factories,''
  Eur.\ Phys.\ J.\ C {\bf 41}, no. 1, 1 (2005)
  doi:10.1140/epjc/s2005-02169-1
  [hep-ph/0406184].

\bibitem{Hocker:2001xe}
  A.~Hocker, H.~Lacker, S.~Laplace and F.~Le Diberder,
  ``A New approach to a global fit of the CKM matrix,''
  Eur.\ Phys.\ J.\ C {\bf 21}, 225 (2001)
  doi:10.1007/s100520100729
  [hep-ph/0104062].

\bibitem{Koike:1971tu}
K.~Koike, M.~Konuma, K.~Kurata and K.~Sugano,
  ``Neutrino production of lepton pairs. 1. -,''
  Prog.\ Theor.\ Phys.\  {\bf 46}, 1150 (1971).
  doi:10.1143/PTP.46.1150

\bibitem{Koike:1971vg}
K.~Koike, M.~Konuma, K.~Kurata and K.~Sugano,
  ``Neutrino production of lepton pairs. 2.,''
  Prog.\ Theor.\ Phys.\  {\bf 46}, 1799 (1971).
  doi:10.1143/PTP.46.1799

\bibitem{Belusevic:1987cw}
R.~Belusevic and J.~Smith,
  ``W - Z Interference in Neutrino - Nucleus Scattering,''
  Phys.\ Rev.\ D {\bf 37}, 2419 (1988).
  doi:10.1103/PhysRevD.37.2419

\bibitem{Brown:1973ih}
R.~W.~Brown, R.~H.~Hobbs, J.~Smith and N.~Stanko,
  ``Intermediate boson. iii. virtual-boson effects in neutrino trident production,''
  Phys.\ Rev.\ D {\bf 6}, 3273 (1972).
  doi:10.1103/PhysRevD.6.3273

\bibitem{CCFR}
S.~R.~Mishra {\it et al.} [CCFR Collaboration],
  ``Neutrino tridents and W Z interference,''
  Phys.\ Rev.\ Lett.\  {\bf 66}, 3117 (1991).
  doi:10.1103/PhysRevLett.66.3117

\bibitem{Aaij:2016flj}
  R.~Aaij {\it et al.} [LHCb Collaboration],
  ``Measurements of the S-wave fraction in $B^{0}\rightarrow K^{+}\pi^{-}\mu^{+}\mu^{-}$ decays and the $B^{0}\rightarrow K^{\ast}(892)^{0}\mu^{+}\mu^{-}$ differential branching fraction,''
  JHEP {\bf 1611}, 047 (2016)
  doi:10.1007/JHEP11(2016)047
  [arXiv:1606.04731 [hep-ex]].

\end{thebibliography}
\end{document}